\documentclass[preprint,12pt]{elsarticle}
\usepackage{epsfig,dsfont}
\usepackage{subfigure}
\usepackage{enumitem}

\journal{Computer Physics Communications}

\begin{document}

\begin{frontmatter}

\title{Surface worm algorithm for abelian
Gauge-Higgs systems on the lattice}

\author{Ydalia Delgado Mercado}
\ead{ydalia.delgado-mercado@uni-graz.at}
\author{Christof Gattringer}
\ead{christof.gattringer@uni-graz.at}
\author{Alexander Schmidt}
\ead{alexander.schmidt@uni-graz.at}
\address{Karl-Franzens University Graz \\
Institute for Physics\\ 
A-8010 Graz, Austria}

\begin{abstract}
The Prokof'ev Svistunov worm algorithm was originally developed for models with nearest
neighbor interactions that in a high temperature expansion are mapped to systems of closed
loops. In this work we present the surface worm algorithm (SWA)  which is a generalization
of the worm algorithm concept to abelian Gauge-Higgs models on a lattice which can be mapped to
systems of surfaces and loops (dual representation). Using Gauge-Higgs models with gauge groups
Z$_3$ and U(1) we compare the SWA  to the conventional approach and to a local update in the
dual representation.  For the Z$_3$ case we also consider finite chemical potential where the
conventional representation has a sign problem which is overcome in the dual representation. 
For a wide range of parameters we find that the SWA clearly outperforms the local update.
\end{abstract}

\begin{keyword}
Lattice QCD, gauge theories, dual representation, Monte carlo methods, worm algorithms.
\end{keyword}

\end{frontmatter}

\setcounter{page}0
\newpage
\section{Introduction}
Monte Carlo simulations are a powerful tool for the analysis of spin systems 
and lattice field theories and Monte Carlo techniques have seen a tremendous
development over the last decades. An important aspect of this development is the choice of the 
representation of a physical system that is optimal for the Monte Carlo simulation. 

A prominent example for the success of a Monte Carlo simulation in an alternative
representation is the Prokof'ev Svistunov worm algorithm \cite{worm}. Originally it
was proposed for the simulation of spin systems in a loop representation. The loop
representation (or dual representation) is obtained from the usual spin language by a high temperature expansion
where the new degrees of freedom are link occupation numbers subject to constraints at the 
sites of the lattice, such that admissible configurations correspond to loops on the
lattice. The worm algorithm not only solves the problem of properly taking into account the
constraints in the Monte Carlo update but turned out to be outperforming many previous
simulation approaches in the conventional formulation \cite{worm_critical_behavior}. 

The worm algorithm concept found many interesting applications also for quantum field theories on a
lattice. In this area a strong motivation for dual representations is the study of quantum
field theories with a chemical potential, where in many cases the standard representation
has complex action and a direct Monte Carlo simulation is not possible. Lattice field
theories that were studied with worm-type algorithms comprise scalar field theories
\cite{scalar,endres,shaileshrev}, fermion systems in various settings, in particular with four fermi terms 
or in the strong coupling limit \cite{fermions}, as well as effective theories for the 
QCD phase diagram \cite{effective}. All these systems have in common that the interaction 
on the lattice is either supported on a single site or on nearest neighbors. The resulting
dual representation thus consists of loops. 

A genuinely new element appears in the dual representation of gauge theories. There the
interaction is based on the plaquettes of the lattice and the corresponding dual variables
(integers assigned to the plaquettes) form surfaces. While for the non-abelian case the 
structure
is rather involved \cite{nonabelian}, abelian gauge theories have a straightforward
representation in terms of closed surfaces. Nevertheless only a few suggestions and 
attempts for a dual simulation of abelian gauge theories can be found in the literature 
\cite{endres,shaileshrev,puregauge} and the main obstacle for a worm-type algorithm is to
efficiently generate the closed surfaces of the dual representation. 

It is interesting to note that the situation is simplified, when matter is coupled
to abelian gauge fields: The dual variables of matter fields are fluxes based on the
links of the lattice that serve as boundaries of the surfaces representing the gauge
degrees of freedom. Despite the fact that an additional field appears, the dual
representation and in particular its Monte Carlo simulation become simpler because the
algorithm now also may use plaquettes bounded by matter flux. A first analysis of a
Gauge-Higgs system in the dual representation with a local Monte Carlo update 
was presented in \cite{dualz3_ref}.

In this article we now present a new Monte Carlo strategy: The surface worm algorithm (SWA) 
which is a generalization of the worm algorithm to a system of surfaces and loops, i.e., dual
representations of abelian Gauge-Higgs models. The SWA uses two main elements: Changing the
flux at an individual link as well as changing a plaquette occupation number and the flux on
two of the links of that plaquette. These steps are used  to efficiently build up filament-like
structures where the link and plaquette occupation numbers are altered. We verify and test  the
surface worm algorithm for lattice Gauge-Higgs models with gauge groups U(1) and Z$_3$. The
latter case has a complex action problem in the conventional approach which is overcome by the
dual representation. We find that in both models  the surface worm algorithm outperforms local
updates. 

\section{Two abelian Gauge-Higgs models and their dual representation}

We use two different Gauge-Higgs models based on the gauge groups Z$_3$ and U(1) to test the surface worm algorithm and explore its properties.
This section defines the two models in their conventional representations and summarizes their dual form in terms of loops of flux and surfaces. For the actual derivation 
of the dual representation we refer to the literature, to \cite{dualz3_ref} for the case of the Z$_3$ model and to \cite{dualu1_ref} for U(1).

\subsection{The Z$_3$ Gauge-Higgs model}
In the conventional form the degrees of freedom of the Z$_3$ Gauge-Higgs model are the gauge
fields $U_{x,\nu} \,$, $\nu = 1,2,3,4$  living on the links of a 4-dimensional lattice, and the
scalar matter fields $\phi_x$ located on the sites. Both sets of degrees of freedom are in the
gauge group   Z$_3 = \{1, e^{i2\pi/3}, e^{-i2\pi/3}\}$.  The lattice we consider has size $V_4 =
N_s^3 \times N_t$ and we use periodic boundary conditions for both fields. The action $S$ is a
sum of the gauge action $S_G$ and the action $S_M$ for the matter fields. The gauge action is
given by
\begin{equation}
S_G\ =\ -\frac{\beta}{2} \sum_x \sum_{\nu < \rho} 
\left[ U_{x,\nu\rho}\ +\ U^*_{x,\nu\rho} \right]\ ,
\label{gauge_action}
\end{equation}
where $U_{x,\nu\rho} = U_{x,\nu}U_{x+\hat{\nu},\rho}U^*_{x+\hat{\rho},\nu}U^*_{x,\rho}$ and $\beta$ is the inverse gauge coupling.
The action for the matter fields is 
\begin{equation}
S_M\; =\; -\kappa \sum_{x,\nu} \left[ 
e^{\mu \delta_{\nu,4}} \phi^*_x \, U_{x,\nu} \, \phi_{x+\hat{\nu}}
\ +\  e^{-\mu \delta_{\nu,4}} \phi_x \, U_{x,\nu}^{\; *} \, \phi_{x+\hat{\nu}}^* \right] \ ,
\end{equation}
where a chemical potential $\mu$ is coupled to the terms in the temporal direction and
the hopping parameter $\kappa$ is a positive real number. The partition sum of the conventional 
representation is given by
$Z = \sum_{\{U,\phi\}} \, e^{-S_G - S_M}$ where the sum  is over all possible field configurations. We stress that in the conventional form the Z$_3$ 
Gauge-Higgs model has a complex action problem at non-zero chemical potential, i.e., $S_M$ is complex for $\mu > 0$.

The partition sum can be rewritten exactly \cite{dualz3_ref} into a dual representation where the new degrees of freedom are link variables $l_{x,\nu} \in \{ -1,0,+1\}$ and plaquette
variables $p_{x,\rho\nu} \in \{ -1,0,+1\}$. The partition function is a sum over all configurations of the link and plaquette variables,
\begin{equation}
Z \; = \; C \, \sum_{\{p,l\}} \; {\cal W} [p,l] \; {\cal C}_S[l] \; {\cal C}_L[p,l] \; .
\label{z3dual}
\end{equation}
The configurations $\{p,l\}$ in  (\ref{z3dual})  come with real and positive weight factors 
\begin{equation}
{\cal W}[p,l]  \; = \;   \bigg(  \prod_{x} \prod_{\nu<\rho } B_\kappa^{\; |p_{x,\nu\rho}|} \bigg) \;  \bigg( \prod_x \prod_{i=1}^3 B_\beta^{\; |l_{x,i}|} \bigg) \; \bigg( \prod_x M_{l_{x,4}} \bigg) \; ,
\label{z3weight}
\end{equation}
with
\begin{equation}
B_\kappa \; = \; \frac{e^{2\kappa}\ -\ e^{-\kappa}}{e^{2\kappa}\ +\ 2e^{-\kappa}}
 \quad , \quad
B_\beta \; = \;  \frac{e^{\beta}\ -\ e^{-\beta/2}}{e^{\beta}\ +\ 2e^{-\beta/2}} \; .
\end{equation}
The overall constant $C$ in  (\ref{z3dual}) is given by $(3 B_\kappa^{\; 3} B_\beta^{\; 6} )^{V_4}$.
The last contribution to the weight (\ref{z3weight}) contains the chemical potential. It is a product over factors $M_{l_{x,4}}$ with 
\begin{equation}
M_l \; = \; \frac{1}{3}\left[ e^{2\kappa \cosh(\mu)}\ +\ 2e^{-\kappa  \cosh(\mu)} 
\cos\left( \kappa\ \sqrt3 \sinh(\mu) - l \frac{2\pi}{3} \right) \right] \; ,
\end{equation}
where $l = +1,0,-1$. Note that the factors $M_l$ are real  and positive also for $\mu > 0$.
Thus the complex action problem is solved in the dual representation.
The configurations $\{p,l\}$ are subjects to the constraints 
\begin{eqnarray}
{\cal C}_L[p,l] & \! = \! & \prod_x \prod_{\nu=1}^4 T\left( \sum_{\rho:\nu<\rho}[p_{x,\nu\rho}
- p_{x-\hat{\rho},\nu\rho}] - \sum_{\rho:\nu>\rho}[ p_{x,\rho\nu}
- p_{x-\hat{\rho},\rho\nu} ] + l_{x,\nu}\right) ,
\nonumber \\
{\cal C}_S[l] & \!=\! & \prod_x T\left( \sum_{\nu=1}^4 [l_{x-\hat{\nu},\nu} - l_{x,\nu}] 
\right) , 
\label{constraint_z3} 
\end{eqnarray}
that both contain the triality function $T(n)$ which is defined to be 1 if $n$ is a multiple of 3
and vanishes otherwise. The constraint ${\cal C}_S[l] $ is a product over sites $x$
of the lattice and enforces that the total flux $\sum_\nu [l_{x-\hat{\nu},\nu} -
l_{x,\nu}]$ at the site $x$ is a multiple of 3. The constraint ${\cal C}_L[p,l] $
is a product over links of the lattice and forces the combined flux from the plaquettes
attached to the link and the  corresponding link variable to be a multiple of 3.

The admissible configurations of the dual variables $p$ and $l$ have the interpretation of surfaces made of non-zero plaquette variables $p_{x,\nu \rho}$. The surfaces can either be closed
(without boundaries) or they are bounded by loops of link variables that compensate the flux at the links that constitute the boundary of the surfaces.

\subsection{The U(1) Gauge-Higgs model} 
\vspace*{3mm}
In the U(1)  Gauge-Higgs model  the degrees of freedom are gauge fields $U_{x,\nu} \in$ U(1) at the links of the lattice
and  a charged scalar Higgs field $\phi_x\ \in\ \mathds{C}$, attached to the sites.  
Again we consider a 4-dimensional lattice with $V_4 = N_s^3 \times N_t$ and periodic boundary conditions for both fields.
The gauge action $S_G$ has the same form as in (\ref{gauge_action}) -- only the link variables are U(1)-valued now. 
The action for the matter fields is given by
\begin{equation}
S_M = \sum_x \left[ \kappa |\phi_x|^2\ +\ \lambda |\phi_x|^4\ -\
\sum_\nu \left( \phi_x^*U_{x,\nu}\phi_{x+\hat{\nu}} + 
\phi_x U^*_{x,\nu}\phi_{x+\hat{\nu}}^* \right) \right]\ .
\label{action_higgs}
\end{equation}
The parameter $\kappa$ denotes $8+m^2$, where $m$ is the bare mass parameter and
$\lambda$ is the quartic coupling.  
The partition sum $Z = \int D[U] D[\phi] e^{-S_G - S_M}$ is given as an integral over 
all field configurations.  

Again the partition sum can be mapped exactly to a dual 
representation. Here we need two sets of link variables, 
$l_{x,\nu} \in \mathds{Z}$, $\overline{l}_{x,\nu} \in \mathds{N}_0$, 
 and plaquette occupation numbers 
$p_{x,\rho\nu} \in \mathds{Z}$. The dual partition function is a sum over all configurations
of the $l$, $\overline{l}$ and $p$ variables,

\begin{equation}
Z = \sum_{\{\overline{l},l\}} \sum_{\{p\}} {\cal W}_M[\overline{l},l] \, {\cal W}_G[p] \,
{\cal C}_S[l] \; {\cal C}_L[p,l] \; .
\label{u1dual}
\end{equation}
The weight factors are 
\begin{eqnarray}
{\cal W}_M[\overline{l},l] &\!\!\! = \!\!\!& \prod_{x,\nu} \frac{1}{(|l_{x,\nu}| 
\! + \! \overline{l}_{x,\nu})! 
\overline{l}_{x,\nu}!} \! \prod_x P\!\left(\! 
\sum_\nu[|l_{x,\nu}| \!+\! |l_{x-\hat{\nu},\nu}| 
+ 2(\overline{l}_{x,\nu} \!+\! \overline{l}_{x-\hat{\nu},\nu})]\! \right)\!, \nonumber \\
{\cal W}_G[p] &\!\!\!=\!\!\! & \prod_{x,\rho<\nu} I_{p_{x,\rho\nu}}( \beta )\ , 
\label{weight_u1} 
\end{eqnarray}
where $I_p(\beta)$ denotes the modified Bessel functions and the $P(n)$ are  
the elementary integrals $P(n) = \int_{0}^{\infty} dx \, x^{n+1} e^{-\kappa x^2 - \lambda x^4}$. 
In a numerical simulation the $P(n)$ are pre-computed and stored for a sufficient number of values
$n$ so they can be used for determining the Metropolis acceptance probabilities efficiently.
Only the $l$ and the $p$ variables are subject to constraints given by
( $\delta(n)$ is here used to denote the Kronecker delta $\delta_{n,0}$ )
\begin{eqnarray}
{\cal C}_L[p,l] & \!=\! &\prod_x \prod_{\nu=1}^4 \delta \left( \, \sum_{\rho:\nu<\rho}[p_{x,\nu\rho}
- p_{x-\hat{\rho},\nu\rho}] - \sum_{\rho:\nu>\rho}[ p_{x,\rho\nu}
- p_{x-\hat{\rho},\rho\nu} ] + l_{x,\nu}\right) ,
\nonumber \\
{\cal C}_S[l] & \!=\! & \prod_x \delta \left( \, \sum_{\nu=1}^4 [l_{x-\hat{\nu},\nu} -
l_{x,\nu}] \right) . 
\label{constraint_u1} 
\end{eqnarray}
The constraints have the same form as for the Z$_3$ case, i.e, we have constraints 
${\cal C}_S[l]$ that are based at the sites for the variables $l$ and constraints
${\cal C}_L[p,l]$ that are based on the links and combine $p$ and $l$ variables. The only
difference is that the triality functions of (\ref{constraint_z3}) are for the U(1) case
replaced by Kronecker deltas, implying that all fluxes must vanish exactly and not only
modulo 3 as in the Z$_3$ case.

\section{Monte Carlo simulation}

In this section we describe the surface worm algorithm (SWA). We also discuss a local Metropolis
algorithm (LMA) for the dual representation which will be used for cross-checking the results from the
SWA. Since the steps used in the SWA may be viewed as a 
decomposition of the local update into smaller elements we first discuss the local update.

For the U(1) Gauge-Higgs model in addition to the plaquette variables $p$ and the constrained
flux variables $l$ we also have the unconstrained link variables $\overline{l}$. Due to the
absence of a constraint we can update the link variables $\overline{l}$ using conventional
Metropolis techniques, which are well documented in textbooks (see, e.g., \cite{landaubinder})
and thus are not discussed in this paper. The update for the constrained variables discussed here is
understood in a background configuration of the $\overline{l}$ variables and in the numerical
tests presented in Section 4 we simply alternate the update of the constrained variables
with sweeps for the $\overline{l}$ fluxes.

\subsection{Local algorithm for the dual representation}

The central aspect of a Monte Carlo simulation in the dual representation is to generate
only admissible configurations, i.e., configurations that obey all constraints. The strategy
which we adopt for the local update is to start from a configuration where all constraints are
obeyed -- typically the configuration where all flux and plaquette variables are set to 0 -- 
and
then to offer local changes of the dual variables that do not violate the constraints. 

\begin{figure}[b!]
\begin{center}
\includegraphics[width=10cm,clip]{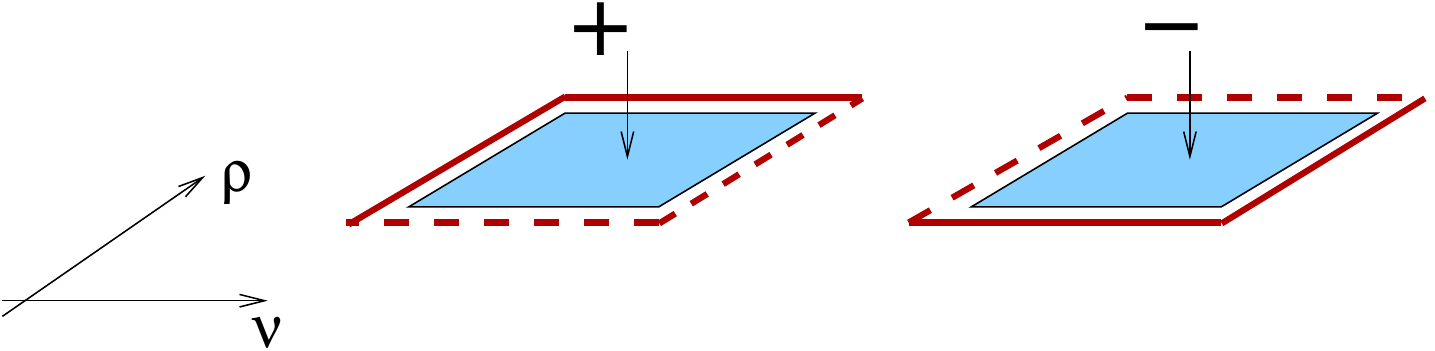}
\end{center}
\vspace{-4mm}
\caption{Plaquette update: A plaquette occupation number is changed by $+1$ (lhs.\ plot) or
$-1$ (rhs.) and the fluxes at the links of the plaquette are changed simultaneously. We use a
full line for an increase by +1 and a dashed line for a decrease by $-1$. 
The directions $1 \le \nu < \rho \le 4$
indicate the plane of the plaquette.}
\label{plaquette}
\end{figure}

The simplest local change is to increase or decrease a plaquette occupation number
$p_{x,\nu\rho}$ by $\pm 1$ and to compensate the violation of the constraint on the
links of the lattice by changing the link fluxes $l_{x,\sigma}$ by $\pm 1$. The two
possible changes (one for increasing $p_{x, \nu \rho}$, one for decreasing) are
illustrated in Fig.~\ref{plaquette}. The change of $p_{x, \nu \rho}$  by $\pm 1$ is
indicated by the signs $+$ or $-$, while for the flux variables we use a dashed line
to indicate a decrease by $-1$ and a full line for an increase by $+1$. It is easy
to see that the pattern of changes for the flux variables not only compensates the
violation of the link-based constraints from changing $p_{x,\nu \rho}$ but also
leaves intact the site-based constraints at all four corners of the plaquette. We
stress that for the case of gauge group Z$_3$ addition of $\pm 1$ is understood
modulo 3, which is the usual addition, except for the cases $1+1 = -1$ and $-1-1 =
1$.

A full sweep of these ``plaquette updates'' consists  
of visiting all plaquettes and offering one of the two changes of Fig.~\ref{plaquette} with
equal probability. The offer is accepted with the usual Metropolis probability 
$\min(1,{\cal W}^{\; \prime}_{loc} / {\cal W}_{loc} )$ where ${\cal W}^{\; \prime}_{loc}$ and   
${\cal W}_{loc}$ are the local weights of the trial configuration and the old configuration.
They can easily be evaluated from the weight factors discussed in the previous section.

\begin{figure}[t!]
\begin{center}
\includegraphics[width=12cm,clip]{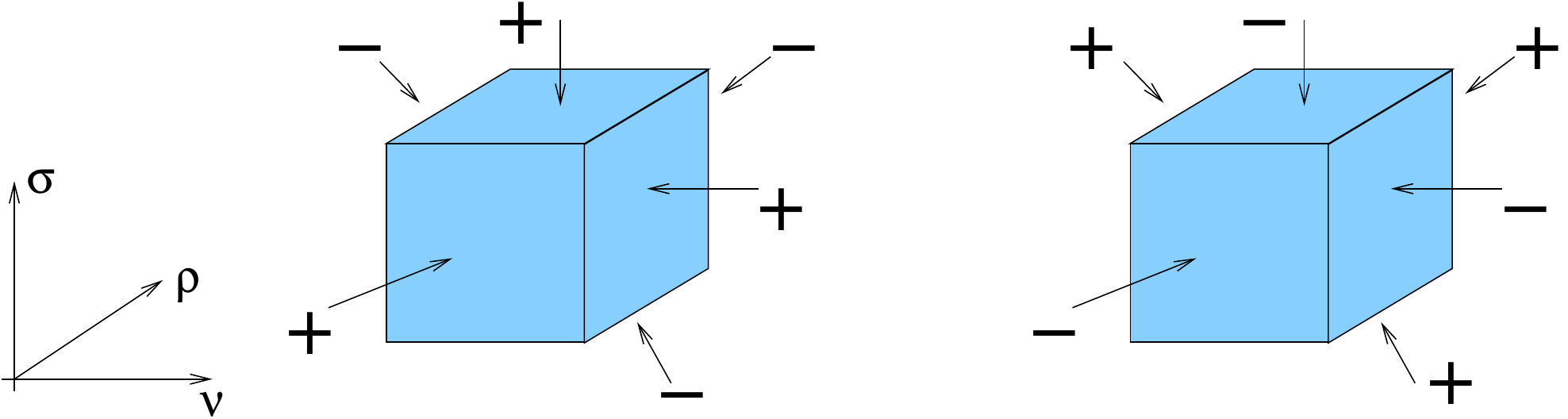}
\end{center}
\vspace{-4mm}
\caption{Cube update: The plaquette occupation numbers of a 3-cube are changed according to the
two patterns we show. The edges of the 3-cube are parallel to 
the directions $1 \leq \nu < \rho < \sigma \leq 4$.}
\label{cubes}
\end{figure}

It is easy to see that the plaquette update alone is ergodic. Nevertheless we found it
advantageous to augment the plaquette update with a ``cube update'' that involves only changes
of plaquette numbers $p_{x,\nu\rho}$. The cube update helps to decorrelate the system in
parameter regions where link flux has a very small Boltzmann weight. The plaquettes on the faces
of 3-cubes of our 4-D lattice are changed according to one of the two patterns shown in 
Fig.~\ref{cubes} (for Z$_3$ addition is again modulo 3). The two possibilities are
offered with equal probability and it is easy to check that the link-based constraints are not
violated, and since no flux variables are involved also the site-based constraints remain
intact.  A full sweep of cube updates consists of visiting all 3-cubes,
offering one of the two changes and accepting them with the Metropolis probability
computed from the local weight factors. 

\subsection{Surface worm algorithm} \label{update}
The surface worm algorithm (SWA) is constructed by breaking up the plaquette update discussed
in the previous subsection into smaller building blocks used to grow filament-like clusters 
on which the flux and plaquette variables are changed. We will first discuss in detail the
SWA  for the Z$_3$ Gauge-Higgs model and then address the  modifications necessary for U(1).

\begin{figure}[t]
\begin{center}
\hspace*{-4mm}
\includegraphics[width=13.5cm,clip]{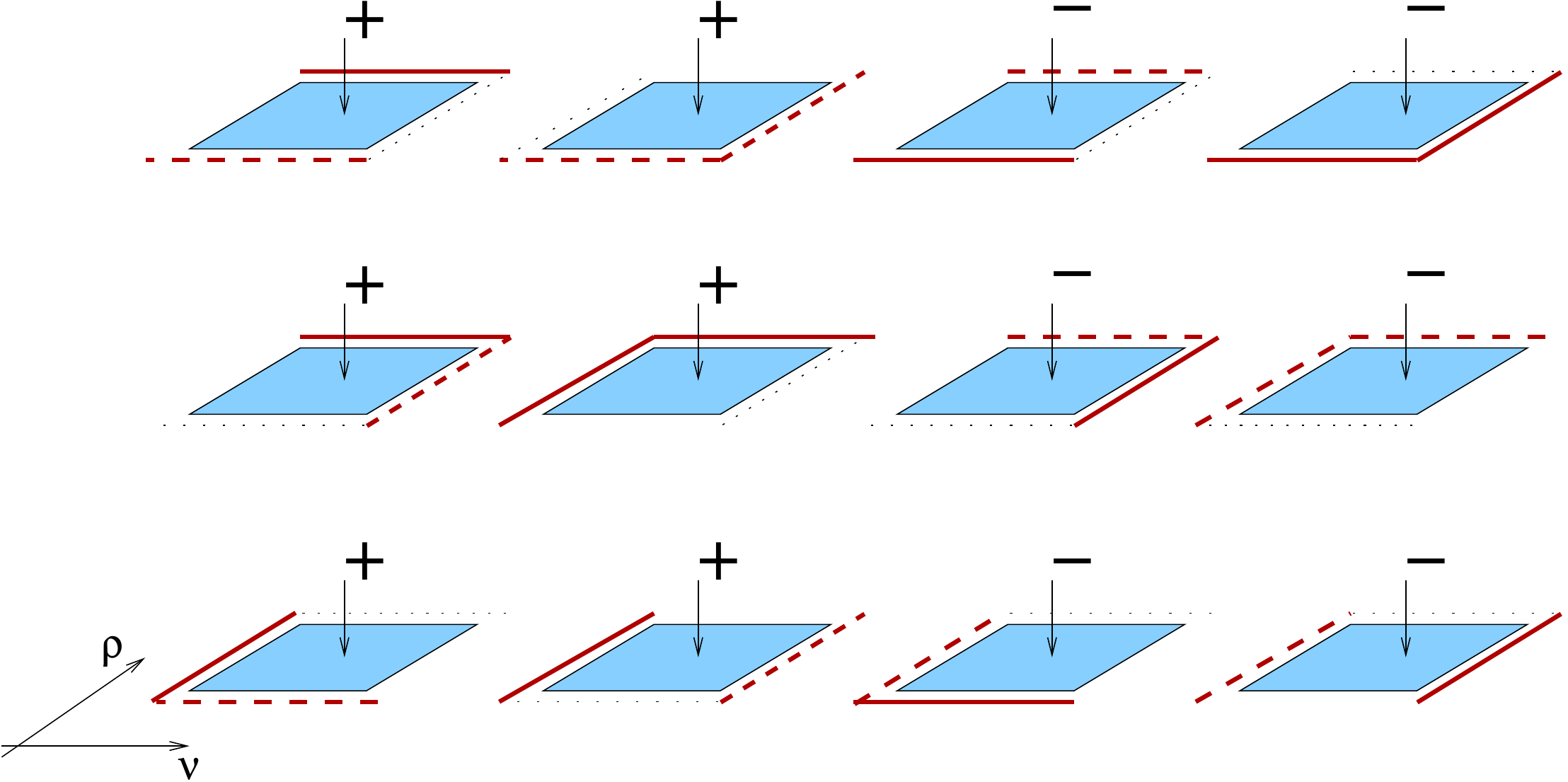}
\end{center}
\vspace{-4mm}
\caption{12 of the 24 possible positive (marked with $+$) and negative segments  in the $\nu$-$\rho$-plane 
($\nu < \rho$).  The remaining 12 segments are exactly the same but
with the position of the empty and dotted links exchanged. Segments in other planes are constructed equivalently.
The plaquette occupation numbers are changed as indicated by the signs. 
The links marked with full (dashed) lines are changed by $+1$ ($-1$). The empty link shows
where the segment is attached to the worm and the dotted link is the new position of the link
$L_V$ where the constraints are violated.  }
\label{segments}
\end{figure}

As for any worm algorithm, in the SWA the constraints   are temporarily violated  at a link
$L_V$ and the two sites at its endpoints. This is done by changing  the flux at a randomly
chosen link  by $\pm 1$ (addition is again modulo 3 for the Z$_3$ case). The defect at $L_V$  is
then propagated through the lattice by offering steps where a plaquette occupation number is changed by $\pm 1$ and
two  flux variables at two of the links of the plaquette. We refer to these structures as
``segments'' and show some examples in Fig.~\ref{segments}. Attaching segments propagates the
link $L_V$ where the constraint is violated through the lattice until the worm decides to
terminate with the insertion of another unit of link flux.  Each step  is accepted
with a Metropolis decision.

Fig.~\ref{segments} shows some examples of  segments that are used by the SWA. The plaquette
occupation numbers are changed by $\pm 1$ as indicated and also the fluxes at two of the links
of the plaquette (again we  use a full line if the flux at a link is increased by $+1$ and
dashed lines for a decrease  by $-1$).   We refer to a segment as a ``positive segment'' if the
plaquette occupation number is increased  (first and second segment shown in
Fig.~\ref{segments}) and use ``negative segment'' otherwise (third and fourth segment).  The empty link
represents the link where a segment is attached to the existing filament-like structure of the SWA and the
dotted link is the new (= shifted) position of the link $L_V$  where the constraints  are
violated (``head of the worm'').

\begin{figure}[p]
\begin{center}
\subfigure[The worm starts by decreasing the flux in $\nu$ direction. Subsequently it adds
a segment in the $\nu$-$\rho$ plane.]{
\includegraphics[width=10.5cm,clip]{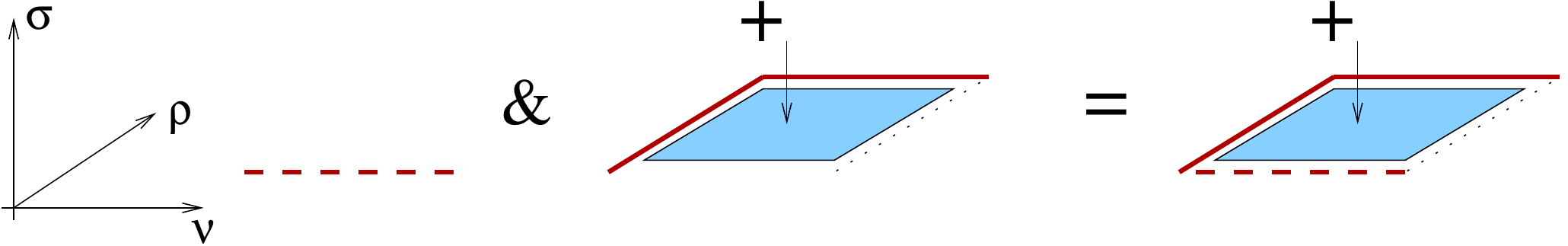}
\label{cube1}
}

\subfigure[The worm adds a segment in the $\rho$-$\sigma$ plane.]{
\includegraphics[width=10.5cm,clip]{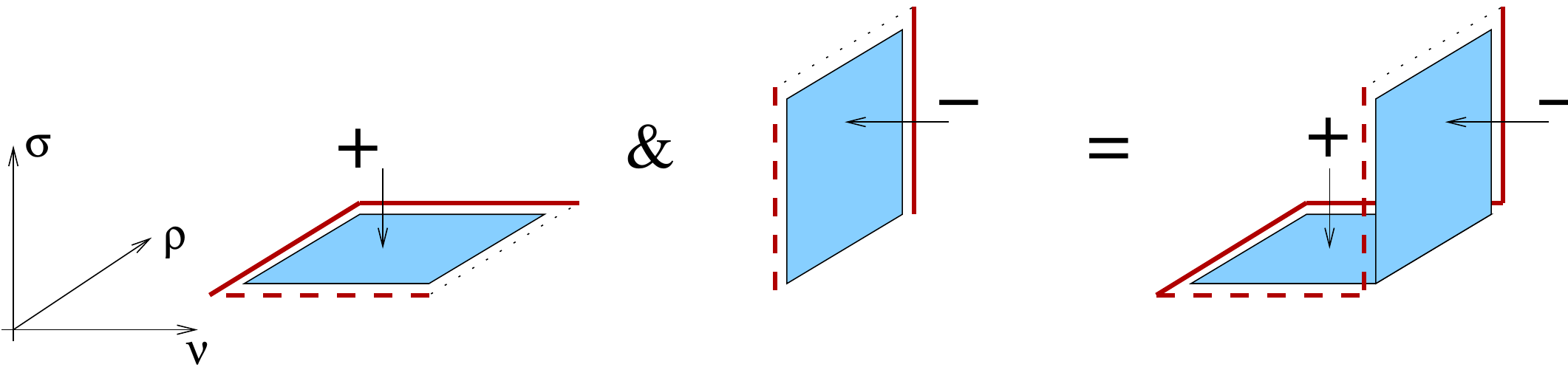}
\label{cube2}
}

\subfigure[The worm adds a segment in the $\nu$-$\rho$ plane.]{
\includegraphics[width=10.5cm,clip]{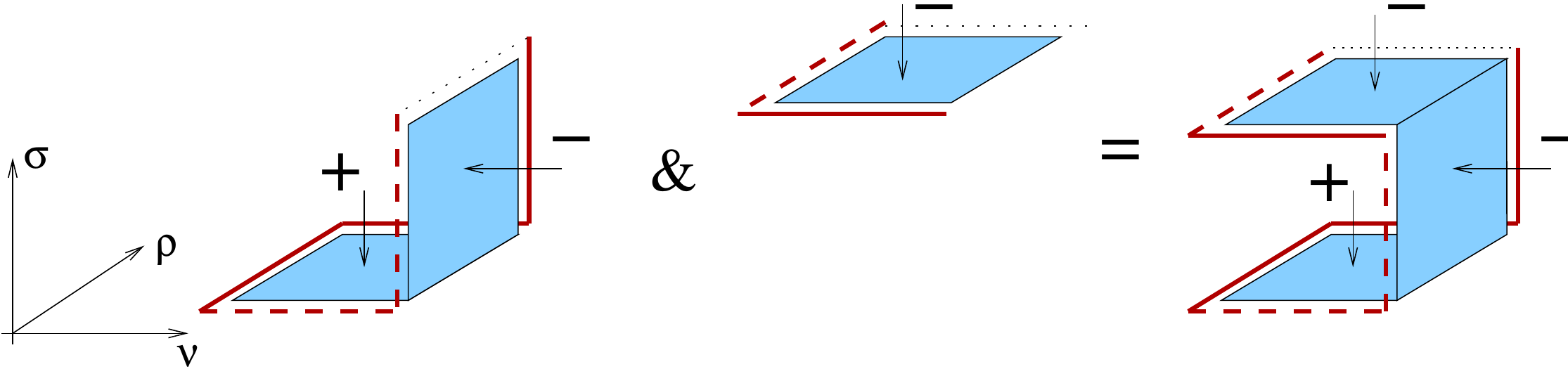}
\label{cube3}
}

\subfigure[A segment in the $\nu$-$\sigma$ plane is added.]{
\includegraphics[width=10.5cm,clip]{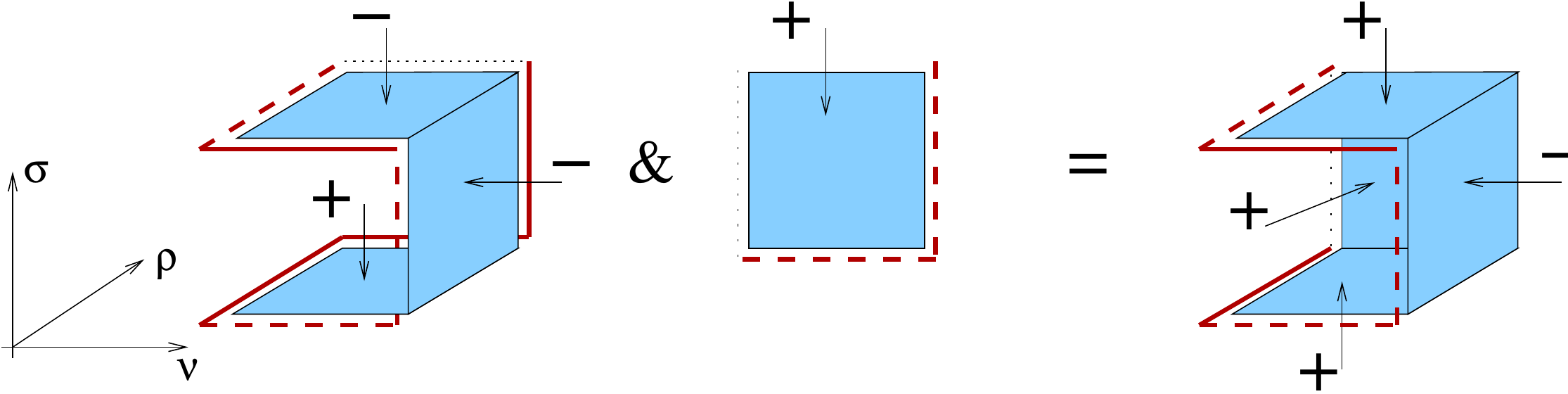}
\label{cube4}
}

\subfigure[The worm decides to saturate the violated constraint by adding a unit of
flux in the $\sigma$ direction and terminates.]{
\includegraphics[width=10.5cm,clip]{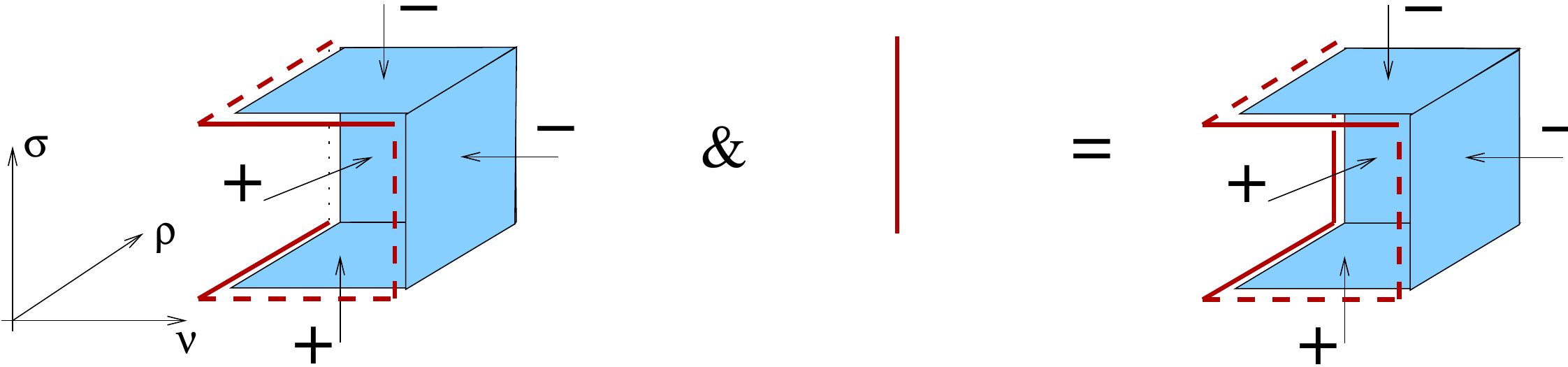}
\label{cube5}
}
\end{center}
\caption{Example of a surface worm algorithm on an initially empty lattice.}
\label{cube}
\end{figure}

Thus the SWA proceeds as follows (for an example see Figs.~\ref{cube1}--\ref{cube5}):

\begin{itemize}
\item The SWA starts at a randomly chosen link $L_0$ 
where the flux is changed by $\pm 1$ (in the example Fig.~\ref{cube1} the
flux is changed by $-1$). At this link
and at its endpoints the constraints are violated, i.e., $L_V = L_0$.

\item Subsequently the SWA either moves $L_V$ by attaching a suitable segment 
(Figs.~\ref{cube1}--\ref{cube4}) or decides to
change the link flux at $L_V$ to heal the violated constraint thus terminating the worm
(Fig.~\ref{cube5}).

\end{itemize}

\noindent
Whenever the worm decides to add a new segment it first randomly determines a new
plane for the segment. This plane has to contain the direction of the link 
$L_V$ that currently violates the constraint. Subsequently the worm 
has to determine whether to insert a positive or a negative 
segment to create only admissible configurations. 
The following steps and Fig.~\ref{signs}
explain how the worm selects an admissible segment 
($1 \leq \nu < \rho < \sigma \leq 4$):

\begin{figure}[t!]
\begin{center}
\subfigure[$L_V \parallel \widehat{\nu}$]{
\includegraphics[width=3.8cm,clip]{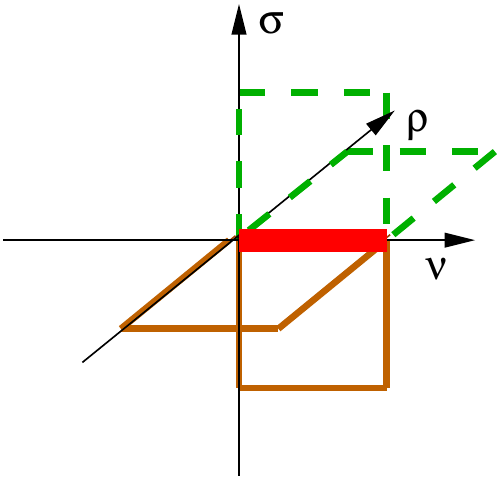}
\label{sign1}
}
\subfigure[$L_V \parallel \widehat{\rho}$]{
\includegraphics[width=3.8cm,clip]{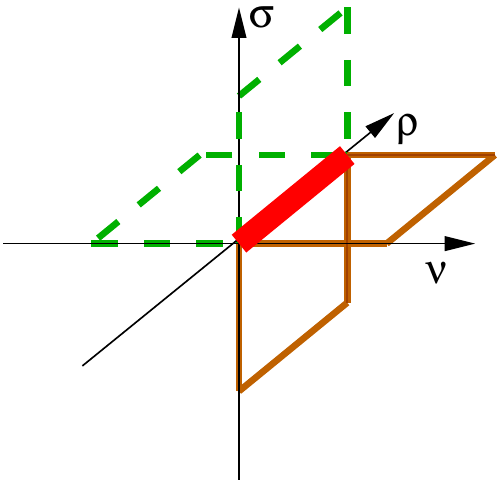}
\label{sign2}
}
\subfigure[$L_V \parallel \widehat{\sigma}$]{
\includegraphics[width=3.8cm,clip]{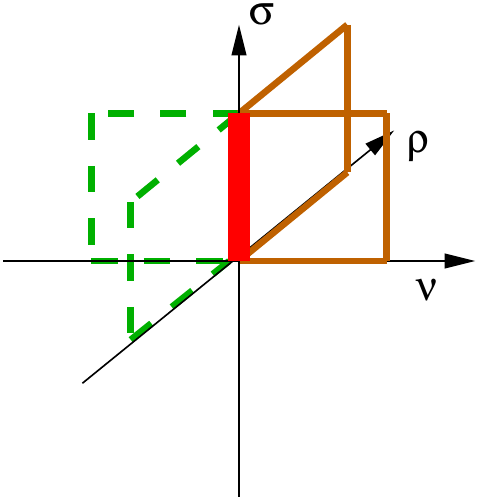}
\label{sign3}
}
\caption{This figure depicts the constraints of the dual partition function.
It can be used to determine whether a positive or negative segment will 
be inserted by the worm: The link $L_V$ where the constraint is violated at the 
current step of the worm either points in $\nu$, $\rho$ or $\sigma$ direction
(plots (a), (b) or (c)), and is marked by a fat link in the corresponding diagrams.
Both the old and the new plaquette are attached to the link and need to be identified
in the corresponding plot.  If they both are surrounded by the same type of line 
(full versus dashed)  the sign of the change of the plaquette variable remains the same,
otherwise an extra factor ($-1$) is taken into account.}
\label{signs}
\end{center}
\end{figure}

\begin{enumerate}
\item Depending on the direction of the link $L_V$ (i.e., $L_V \parallel \widehat{\nu}$, 
$L_V \parallel \widehat{\rho}$ or $L_V \parallel \widehat{\sigma}$) identify $L_V$ 
as the central link surrounded by four plaquettes in one of the diagrams of Fig.~\ref{signs}.
\item Identify the ``old plane'' and the ``new plane'':
      \\Old plane: plane of the last successfully updated segment.
      \\New plane: plane of the new trial segment.
\item If the plaquettes in the old and the new plane are marked by different lines (full
versus dashed) keep the same type of segment.
      Otherwise change the type of segment from positive to negative or vice-versa.
\end{enumerate}

\noindent
Note that when the worm attempts to revisit the last updated plaquette
(i.e., it moves backwards) then the new and old planes coincide. 
Thus the segment changes  and the last move of the worm is undone. 

\begin{figure}[t!]
\begin{center}
\subfigure[
The worm starts by increasing the flux in the $\nu$-direction and then adds a segment in the 
$\nu$-$\rho$ plane.]{
\includegraphics[width=9.7cm,clip]{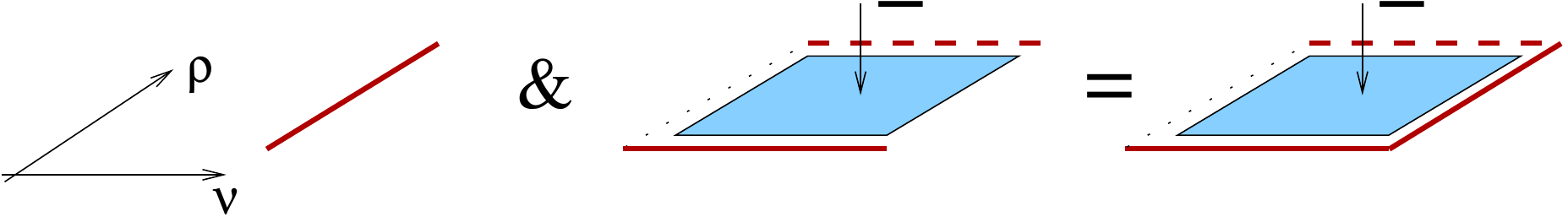}
\label{plaq1}
}
\newline
\subfigure[The worm decides to saturate the violated constraint by decreasing the flux at
$L_V$ by one unit and terminates.]{
\hspace*{-10mm}
\includegraphics[width=9.8cm,clip]{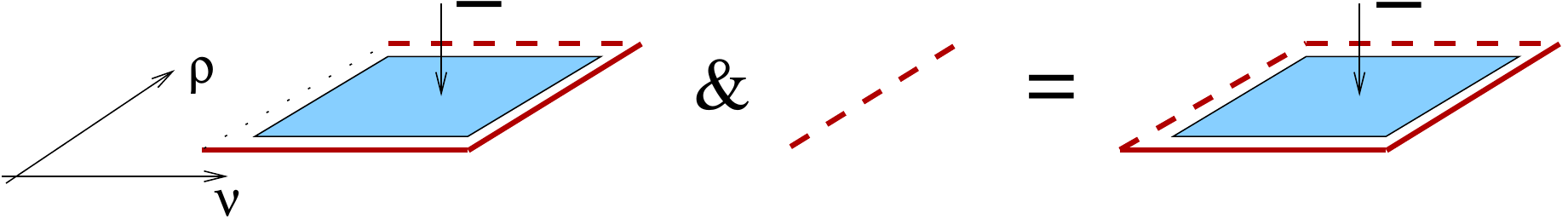}
\label{plaq2}
}
\end{center}
\vspace*{-4mm}
\caption{Example how the worm generates the local plaquette update discussed in the previous
subsection.}
\label{plaq}
\end{figure}

In addition to the example of Fig.~\ref{cube}, in Fig.~\ref{plaq} we show a short 
worm that generates the plaquette update of the local algorithm discussed in the previous
subsection. We have already stressed that the plaquette update is ergodic and the SWA thus is
ergodic too.

The pseudo-code listed below describes the algorithm. For the coordinates of
plaquettes we use $P$, and $L$ for the coordinates of links. In particular the 
link where the constraints are violated (head of the
worm) is denoted by $L_V$. By ${s}_P =
(p_P,l_0,l_1)$  we denote the current occupation numbers of a segment,
i.e., the occupation number $p_P$ of the plaquette at
$P$ and the two links fluxes $l_0$ and $l_1$ which are changed in the type of
segment chosen (in the examples of segments shown in Fig.~\ref{segments} $l_0$ and
$l_1$ are the link fluxes marked by full or dashed lines). The variable 
$\Delta_{{s}} = (\delta_p,\delta_{l_0},\delta_{l_1})$ 
denotes the change of the occupation numbers of ${s}_P$. Note that the 
sign of the change $\delta_p$  (``positive segment'' versus ``negative segment'')
has to be
chosen according to the rules stated in the discussion of Fig.~\ref{signs}.
By
$x \oplus y$ we denote the addition modulo 3 which is the usual addition operation 
except in the cases $+1\oplus+1=-1$ and $-1\oplus-1=+1$. By {\tt weight\_ratio}$(b
\leftarrow a)$ we denote the ratio ${\cal W}_{loc}^{\; \prime} / {\cal W}_{loc}$ when
changing an element $a$ into $b$. Here $a$ and $b$ are either a link flux before and
after the change by $\pm1$ or a full segment (a plaquette number $p_P$ and
two link fluxes $l_0, l_1$) before and after the respective changes. Finally,  {\tt rand()} is a random
number generator for uniformly distributed real numbers in the interval $[0,1)$.

\vskip5mm
\noindent
\underline{{\bf Pseudocode for surface worms:}}
\vskip3mm
{\tt 
\noindent select a lattice link $L_0$ randomly

\noindent select $\delta_l \in \{-1,+1\}$ randomly

\noindent $l' \, \longleftarrow \, l_L \, \oplus \, \delta_l$

\noindent if rand() $ \leq \; $ weight\_ratio($l' \, \leftarrow \, l_L$)

\hspace*{5mm} $l_L \, \longleftarrow \, l'$

\hspace*{5mm} $L_V \, \longleftarrow \,$ $L_0$

\noindent else

\hspace*{5mm} terminate worm

\noindent end if

\vskip3mm
\noindent repeat until worm is complete:

\noindent select a direction $\rho \in \{\pm \hat{1}, \pm \hat{2},\pm \hat{3},\pm \hat{4}\}$

\noindent if $\widehat{\rho} \parallel L_V$ then

\hspace*{5mm} select $\delta_l$ such that the violated constraint at $L_V$ is healed

\hspace*{5mm} $l' \, \longleftarrow \, l_{L_V} \, \oplus \, \delta_l$

\hspace*{5mm} if rand() $ \leq \; $ weight\_ratio($l' \, \leftarrow \, l_{L_V}$)

\hspace*{10mm} $l_{L_V} \, \longleftarrow \, l'$

\hspace*{10mm} terminate worm

\hspace*{5mm} end if

\noindent else

\hspace*{5mm} the plaquette $P$ for a new segment is spanned by $L_V$ and $\widehat{\rho}$
 
\hspace*{5mm} randomly select $L^\prime_V \neq L_V$ from the links bounding $P$

\hspace*{5mm} choose $\Delta_{{s}}$ such that the constraint at $L_V$ is healed

\hspace*{5mm} ${s}' \, \longleftarrow \, {s}_{P} \, \oplus \, \Delta_{{s}}$

\hspace*{5mm} if rand() $ \leq \;$ weight\_ratio(${s}' \, \leftarrow \, {s}_{P}$)

\hspace*{10mm} ${s}_{P} \, \longleftarrow \, {s}'$

\hspace*{10mm} $L_V \, \longleftarrow L_V^\prime$

\hspace*{5mm} end if

\noindent end repeat until worm is complete
}

\vskip5mm
\noindent
It is straightforward to show detailed balance using the Boltzmann 
weights and that the algorithm is ergodic.

 \newpage

\vskip5mm

\noindent
\underline{\bf Modifications for the U(1) gauge-Higgs surface worm algorithm}

\vskip3mm
\noindent
From Eq.(\ref{z3dual}) and Eq.(\ref{u1dual})  we observe 
that the SWA has to be adapted in order to simulate the U(1) model:

\begin{itemize}[leftmargin=0.5cm]
\item Due to the extra unconstrained set of link variables 
$\overline{l}_{x,\nu}$, for U(1) a full sweep consists of
a worm sweep ($V_4 = N_s^3 N_t$ worms) to update the $l_{x,\nu}$ and $p_{x,\nu\rho}$ 
plus a conventional local Metropolis sweep to update all $\overline{l}_{x,\nu}$.

\item To extend the range of the constrained variables to all
integer numbers and enforce the total flux at every link and site 
to vanish, the operation $x \oplus y$ is replaced by a normal 
addition $x + y$.  In the pseudo-code: 
${s}_P \, \oplus \, \Delta_{{s}}$ 
is replaced by ${s}_{P}\ +\ \Delta_{{s}}$.
\end{itemize}

\section{Assessment of the surface worm algorithm}

\subsection{Validity of the SWA}
To evaluate the validity of the algorithm 
we will use several thermodynamical 
observables and their susceptibilities.
For both models we study the first and second derivatives with respect to the inverse gauge coupling $\beta$,
i.e., the plaquette expectation value and its susceptibility,

\begin{equation}
\langle U \rangle = \frac{1}{6 N_s^3 N_t}\frac{\partial}{\partial \beta} \ln\ Z\quad , \quad
\chi_{U} = \frac{1}{6 N_s^3 N_t}\frac{\partial^2}{\partial \beta^2} \ln\ Z\ .
\end{equation}

\noindent For the Z$_3$ case we also consider the particle number density $n$ 
and its susceptibility which are the derivatives 
with respect to the chemical potential,

\begin{equation}
n  = \frac{1}{N_s^3 N_t}\frac{\partial}{\partial \mu} \ln\ Z\quad , \quad
\chi_{n} = \frac{1}{N_s^3 N_t}\frac{\partial^2}{\partial \mu^2} \ln\ Z\ .
\end{equation}

\noindent Finally, for the U(1) model we analyze the derivatives with 
respect to $\kappa$,

\begin{equation}
\langle |\phi|^2 \rangle = \frac{1}{N_s^3 N_t}\frac{\partial}{\partial \kappa} \ln\ Z\quad , \quad
\chi_{|\phi|^2} = \frac{1}{N_s^3 N_t}\frac{\partial^2}{\partial \kappa^2} \ln\ Z\ .
\end{equation}

\vskip3mm
The correctness of the flux representation has 
already been established in \cite{dualz3_ref,dualu1_ref}.  
Thus here we can focus on the SWA. To check for correctness we 
compare the SWA results to the data coming from the local Metropolis algorithm 
(LMA) in the flux representation and for the cases where there is no sign problem 
also to results from a conventional approach in the standard representation.

For all simulations we used thermalization and decorrelation sweeps 
(see below for their numbers). For the SWA one sweep consists of $V_4 = N_s^3 N_t$ worms and for 
the case of U(1) also of a sweep through all unconstrained link variables
$\overline{l}_{x,\nu}$. For the LMA a sweep is defined as a sequence of plaquette 
updates for all
$6V_4$ plaquettes plus cube updates for all $4V_4$ cubes. For the U(1) model and the Z$_3$ case at $\mu =
0$ we can also compare to the conventional approach where as usual a sweep is defined 
as applying one
local Metropolis update to all degrees of freedom.    
All error bars we show were determined using a Jackknife analysis and are corrected 
with the factors from the respective autocorrelation times (see below).

\begin{figure}[t!]
\begin{center}
\includegraphics[width=12.5cm,clip]{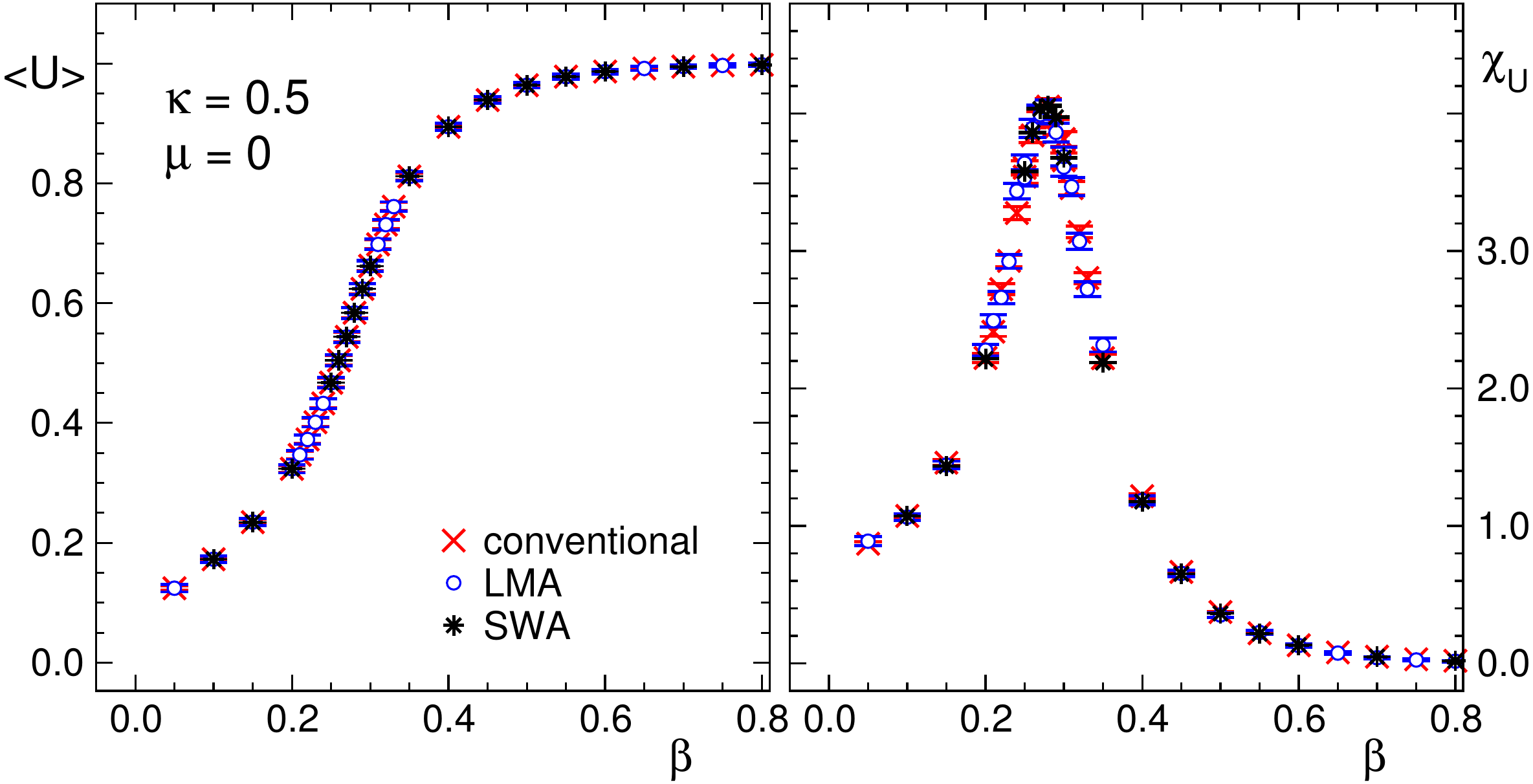}
\end{center}
\vspace{-4mm}
\caption{Z$_3$ model: $\langle U \rangle$ and $\chi_U$ at $\kappa=0.5$ and $\mu = 0$ as a function of 
$\beta$ on a 
$10^4$ lattice. We compare the results of the SWA (asterisks) to the LMA (circles) and the
conventional approach (crosses).}
\label{mug05}
\end{figure}

\begin{figure}[t!]
\begin{center}
\subfigure{
\hspace*{-3.5mm}
\includegraphics[width=127.5mm,clip]{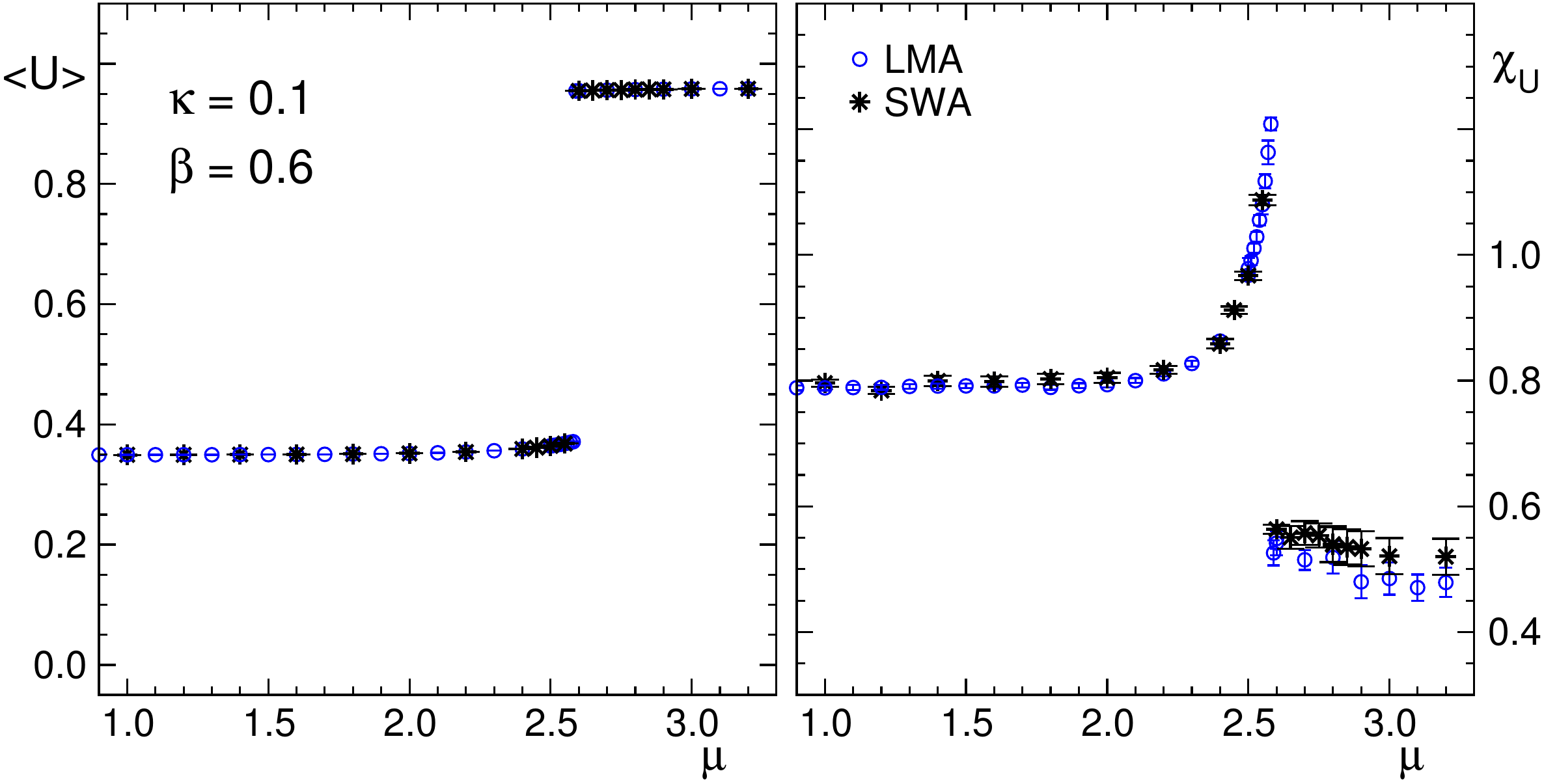}
\label{b06u}
}
\subfigure{
\includegraphics[width=126mm,clip]{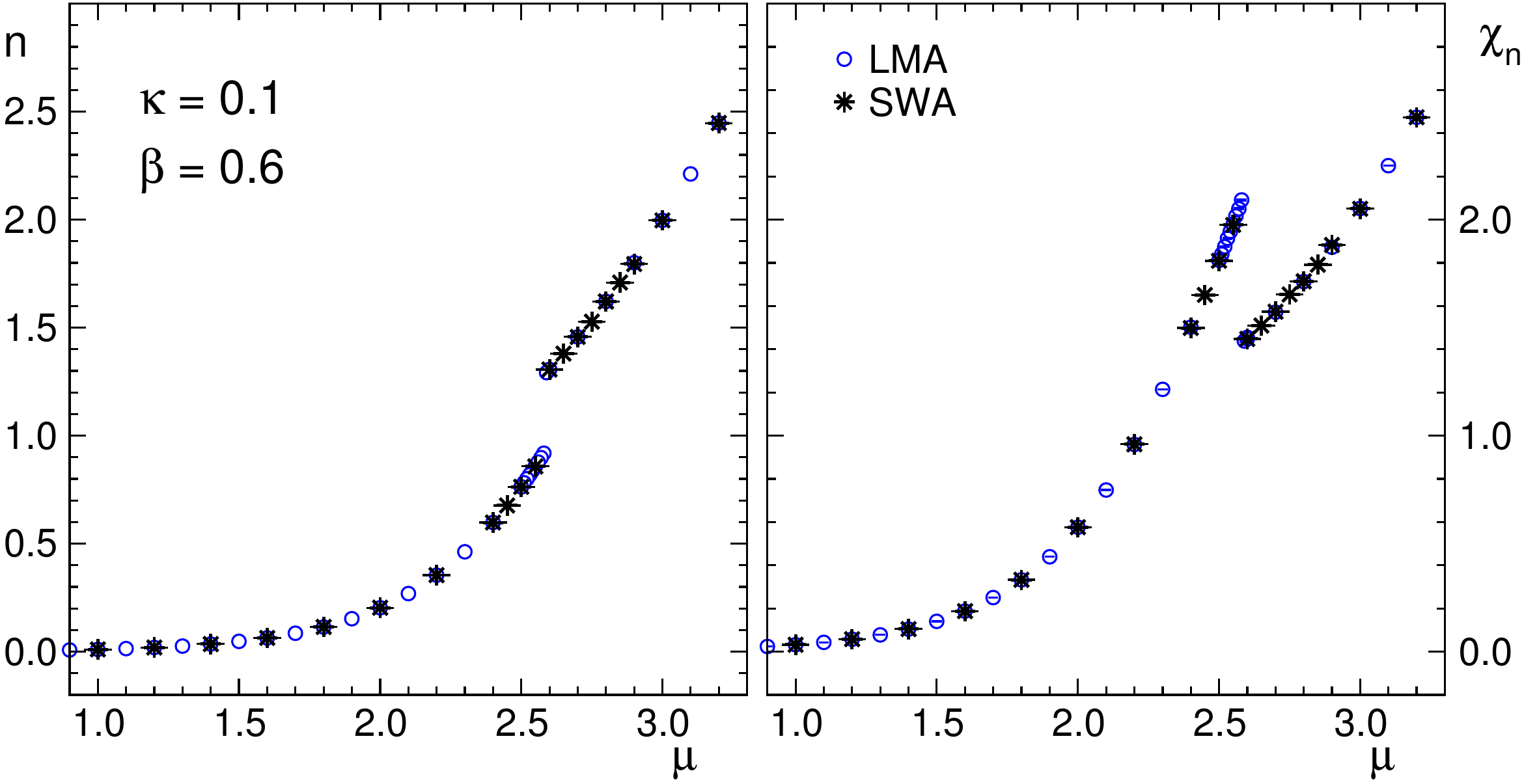}
\label{b06n}
}
\end{center}
\vspace{-4mm}
\caption{Z$_3$ model: The observables $\langle U \rangle$, $\chi_U$, $n$ and $\chi_n$ 
as a function
of $\mu$ at $\kappa=0.1$ and $\beta = 0.6$ on a $8^3 \times 50$ lattice. We compare the
results from the SWA (asterisks) and the LMA (circles).}
\label{b06}
\end{figure}

For the Z$_3$ model we compared simulations 
for several parameter sets and found very good agreement
of the results from the different approaches. As examples we show results for two 
parameter sets: 1) The behavior across a crossover transition as a function of $\beta$ at
$\kappa = 0.5$ and $\mu = 0$ (no complex action problem)
on a $10^4$ lattice (Fig.~\ref{mug05}).  
2) The behavior across a first order transition as a function of $\mu$ at 
$\kappa = 0.1$ and $\beta = 0.6$ on a $8^3 \times 50$ lattice (Fig.~\ref{b06}).
In the latter case the standard representation has a complex action problem and we only can compare
the results from SWA and LMA. For both tests we used $10^6$ equilibration sweeps and $10^6$ 
measurements separated by $10$ sweeps for decorrelation.

\begin{figure}[t!]
\begin{center}
\subfigure{
\hspace*{-3.3mm}
\includegraphics[width=124.5mm,clip]{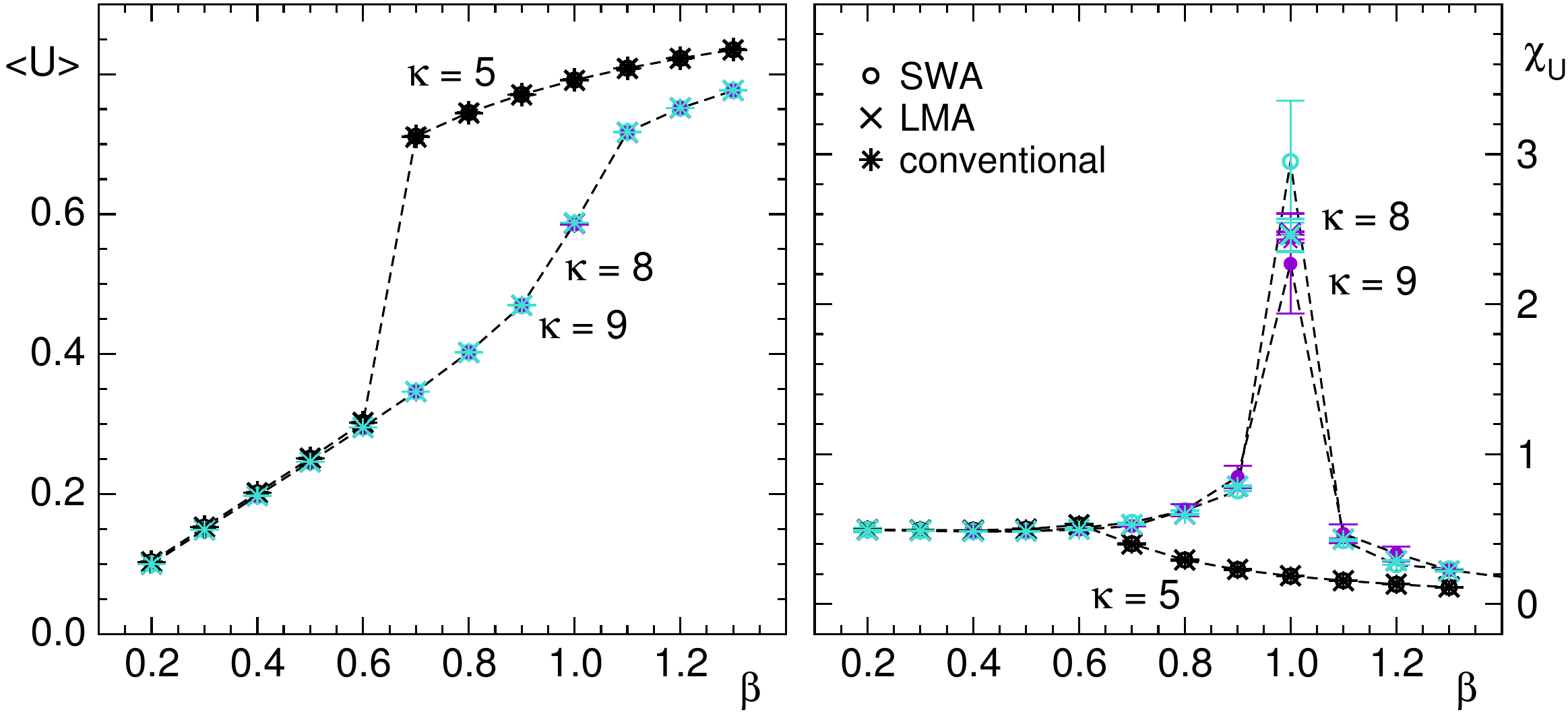}
\label{results1_u1_0}
}

\subfigure{
\hspace*{-5mm}
\includegraphics[width=129mm,clip]{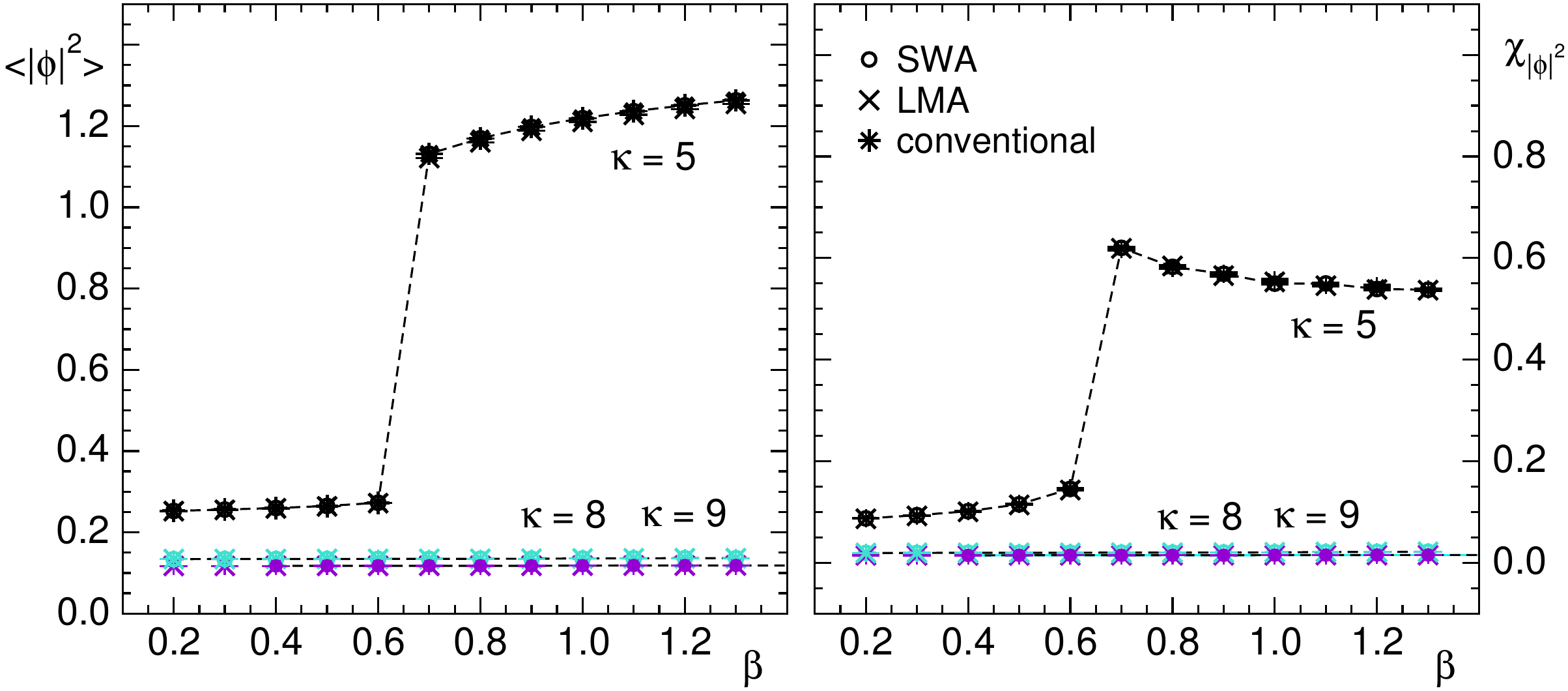}
\label{results1_u1_1}
}
\end{center}
\vspace{-4mm}
\caption{U(1) model: Observables as a function
of $\beta$ at $\lambda = 1.0$ for $\kappa = 5,8$ and 9 on a $10^4$ lattice.  We compare results from three
algorithms: The conventional approach (asterisks), the SWA (circles) and
the LMA (crosses).}
\label{results1_u1}
\end{figure}

Similarly we also confirmed the correctness of the SWA in the U(1) model 
checking the agreement of all three approaches 
at different parameters and lattice sizes.  
As an example, Fig.~\ref{results1_u1} shows the results obtained with 
the LMA (crosses), with the SWA (circles) and the conventional 
approach (asterisks)
at $\lambda = 1$ and $\kappa =$ $5$, $8$  and $9$ 
on a $10^4$ lattice. For this test we used $10^5$ equilibration
sweeps and $10^5$ measurements separated by $10$ sweeps for decorrelation.
As for the Z$_3$ case we find very good agreement among the different
approaches thus establishing the correctness of the SWA also for the U(1) model. 

\subsection{Characteristic quantities of the algorithms}
For a meaningful comparison of the performance and autocorrelation times of the SWA and LMA
algorithms we study suitable characteristic quantities in order
to describe the behavior of both algorithms in 
different regions of the parameter space. The definitions are patterned after related
quantities introduced for the analysis of worm algorithms with open ends \cite{effective}.

\begin{itemize}

\item Plaquette changes ${\cal P}$:
\begin{eqnarray}
{\cal P}  =  \mbox{average number of plaquettes changed per update}
\nonumber 
\end{eqnarray}

\item Starting fraction ${\cal S}$:
\begin{eqnarray}
{\cal S} = \frac{\mbox{number of successful update starts}}{
  \mbox{number of all start attempts}} \; \leq \; 1
\nonumber  
\end{eqnarray}

\item Cost ratio ${\cal C}$:
\begin{eqnarray}
{\cal C} = \frac{\mbox{number of attempted changes (plaquettes and links)}}{
\mbox{number of accepted changes}} \; \geq \; 1
\nonumber 
\end{eqnarray}

\end{itemize}
In these definitions ''update'' refers to one surface worm for the SWA case. For the LMA
it is the average of a plaquette and a cube update which we consider in a mix of 6$V_4$ plaquette
updates and 4$V_4$ cube updates per LMA sweep (see above). From the definition of these characteristic quantities
it is obvious that an optimal algorithm is
characterized by a large value of ${\cal P}$ and values of ${\cal S}$ and ${\cal R}$ close to
1.

In Table \ref{count_z3} we show the characteristic quantities for the SWA and LMA algorithms
in the Z$_3$ case. We compare three different sets 
of parameters denoted by {\it Z-1, Z-2} and {\it Z-3} (see the first column for the corresponding
parameter values) and four different volumes (second column).  The parameters of {\it Z-1} are located
below the condensation transition shown in Fig.~\ref{b06}, the set {\it Z-2} is in the condensed
phase (compare Fig. 11 from \cite{dualz3_ref}) and
the set {\it Z-3} is inside the crossover region of Fig.~\ref{mug05}. 

Table \ref{count_z3} demonstrates that the SWA has a larger probability for starting
an update than the LMA  (${\cal S}_{SWA} > {\cal S}_{LMA}$ for all data sets and
volumes). Furthermore the cost ratio ${\cal R}$ of the SWA  is smaller or equal
(equal only for the set {\it Z-2}) to the LMA case.  These two quantities indicate
that the SWA is more effective than the LMA. The observation that ${\cal
P}$ is larger for the LMA is mainly due to the fact that an accepted cube update  of
the LMA changes 6 plaquettes (although at the cost of a low acceptance rate). It is
interesting to note that the values for the  characteristic quantities are
essentially independent of the volume.

Table \ref{count_u1} collects the data for the U(1) case. 
Here we consider three different sets of parameters {\it U-1, U-2, U-3} (first column) 
on four different volumes
(second column). The set {\it U-2} is located very close
to the transition shown in Fig.~\ref{results1_u1}, the set {\it U-1} is below and 
the set {\it U-3} above the transition.

The general behavior for the characteristic quantities is essentially the same as in the Z$_3$ case: For all sets the 
starting probability of the SWA is larger than that of the LMA, and also the cost efficiency is considerably better for the SWA.
As in the Z$_3$ case we find that the average number of updated plaquettes ${\cal P}$ is larger for the LMA, which also here is due to the cube updates,
which, however, have a much lower acceptance rate as is obvious from ${\cal S}$ and ${\cal C}$. 
The difference in the characteristic quantities between the $4^4$ and larger volumes
for the set {\it U-2} is due to finite-size effects:  In the smallest volume the transition
is rounded and slightly shifted towards smaller values of
$\beta$, such that for the smallest volume the parameters we work at are 
further remote 
from the transition and both algorithms are more efficient.

\begin{table}[t!]
\begin{center}
\hspace*{-1mm}
\begin{tabular}{l|r|lll|lll}
\hline
Parameters & $V$ \hspace{3mm} & ${\cal S}_{SWA}$  & ${\cal P}_{SWA}$  & ${\cal C}_{SWA}$  & ${\cal S}_{LMA}$  & ${\cal P}_{LMA}$  & ${\cal
C}_{LMA}$\\
\hline
{\bf Set:} {\it Z-1}      & $ 4^3\times 50$ & 0.203 & 0.095 & 6.892 & 2.9e-3 & 4.456 & 320.7 \\
$\kappa = 0.1,$    & $ 8^3\times 50$ & 0.203 & 0.095 & 6.892 & 2.9e-3 & 4.455 & 320.7 \\
$\beta = 0.6,$     & $12^3\times 50$ & 0.203 & 0.095 & 6.892 & 2.9e-3 & 4.455 & 320.7 \\
$\mu = 2.0$        & $16^3\times 50$ & 0.203 & 0.095 & 6.892 & 2.9e-3 & 4.455 & 320.7 \\
\hline
{\bf Set:} {\it Z-2}      & $ 4^3\times 50$ & 0.245 & 1.196 & 5.319 & 0.172 & 5.384 & 5.346 \\
$\kappa = 0.1,$    & $ 8^3\times 50$ & 0.244 & 1.186 & 5.431 & 0.172 & 5.384 & 5.346 \\
$\beta = 0.8,$     & $12^3\times 50$ & 0.245 & 1.199 & 5.320 & 0.172 & 5.384 & 5.346  \\
$\mu = 1.6$        & $16^3\times 50$ & 0.244 & 1.187 & 5.425 & 0.172 & 5.384 & 5.346 \\
\hline
{\bf Set:} {\it Z-3}      & $ 4^4$ \hspace{3mm} & 0.697 & 0.802 & 3.081 & 0.098 & 1.286 & 10.88 \\
$\kappa = 0.5,$    & $ 8^4$ \hspace{3mm} & 0.698 & 0.802 & 3.081 & 0.098 & 1.286 & 10.88 \\
$\beta = 0.28,$    & $12^4$ \hspace{3mm} & 0.698 & 0.802 & 3.081 & 0.098 & 1.286 & 10.88 \\
$\mu = 0.0$        & $16^4$ \hspace{3mm} & 0.697 & 0.802 & 3.081 & 0.098 & 1.286 & 10.88 \\
\hline
\end{tabular}
\end{center}
\caption{Characteristic quantities for the Z$_3$ model
(see the text for their definitions).  
We used $10^6$ steps for equilibration and $10^6$ measurements
separated by $2$ steps for decorrelation. The errors are smaller 
than the last digit we show.}
\label{count_z3}
\end{table}
 
\begin{table}[h]
\begin{center}
\hspace*{-1mm}
\begin{tabular}{l|c|lll|lll}
\hline
Parameters & \hspace{1mm} $V$ \hspace{1mm} & ${\cal S}_{SWA}$  & ${\cal P}_{SWA}$  & ${\cal C}_{SWA}$  & ${\cal S}_{LMA}$  & ${\cal P}_{LMA}$  & ${\cal
C}_{LMA}$\\
\hline
{\bf Set :} {\it U-1}    & $ 4^4$ & 0.201 & 0.085 & 6.899 & 1.2e-3 & 1.277 & 904.6 \\
$\kappa = 5,$    & $ 8^4$ & 0.201 & 0.085 & 6.902 & 1.2e-3 & 1.278 & 909.2 \\
$\lambda = 1,$   & $12^4$ & 0.201 & 0.085 & 6.902 & 1.2e-3 & 1.278 & 909.4 \\
$\beta = 0.40$   & $16^4$ & 0.201 & 0.085 & 6.902 & 1.2e-3 & 1.278 & 909.4 \\
\hline
{\bf Set:} {\it U-2}    & $ 4^4$ & 0.681 & 1.275 & 3.310 & 0.167  & 1.813 & 6.263 \\
$\kappa = 5,$    & $ 8^4$ & 0.220 & 0.199 & 6.124 & 4.6e-3 & 2.243 & 224.3 \\
$\lambda = 1,$   & $12^4$ & 0.220 & 0.198 & 6.124 & 4.6e-3 & 2.243 & 224.3 \\
$\beta = 0.65$   & $16^4$ & 0.220 & 0.198 & 6.124 & 4.6e-3 & 2.243 & 224.3 \\
\hline
{\bf Set:} {\it U-3}    & $ 4^4$ & 0.107 & 0.100 & 8.775 & 0.061 & 5.962 & 14.82 \\
$\kappa = 8,$    & $ 8^4$ & 0.107 & 0.100 & 8.773 & 0.061 & 5.962 & 14.92 \\
$\lambda = 1,$   & $12^4$ & 0.107 & 0.100 & 8.774 & 0.060 & 5.962 & 14.91 \\
$\beta = 1.10$   & $16^4$ & 0.107 & 0.101 & 8.766 & 0.060 & 5.962 & 14.91 \\
\hline
\end{tabular}
\end{center}
\caption{Characteristic quantities for the U(1) model 
(see the text for their definitions).  
We used $10^6$ steps for equilibration and $10^6$ measurements
separated by $2$ steps for decorrelation.
The errors are smaller than the last digit we show.}
\label{count_u1}
\end{table}

\begin{figure}[h!]
\begin{center}

\includegraphics[width=0.8\textwidth,clip]{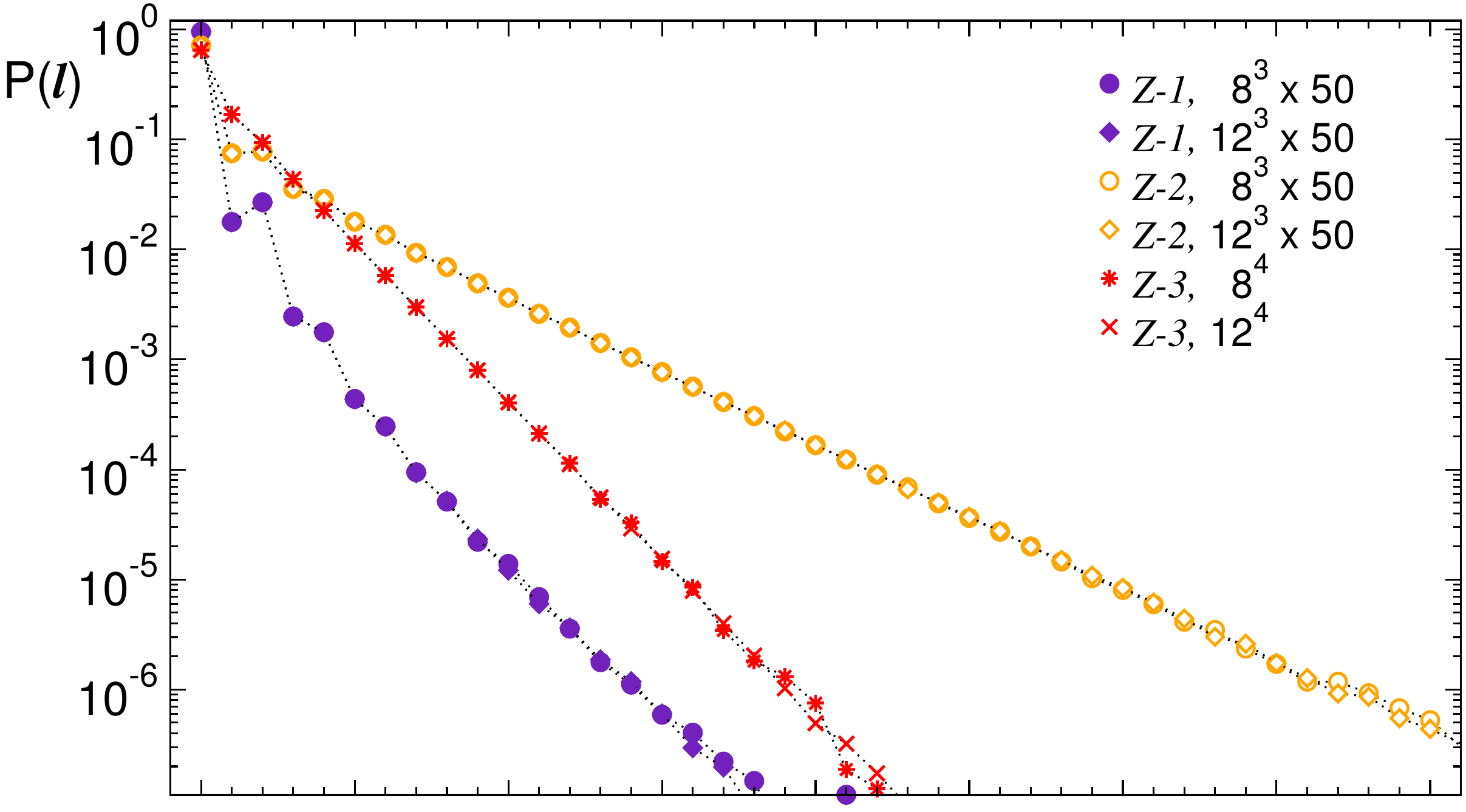}
\includegraphics[width=0.8\textwidth,clip]{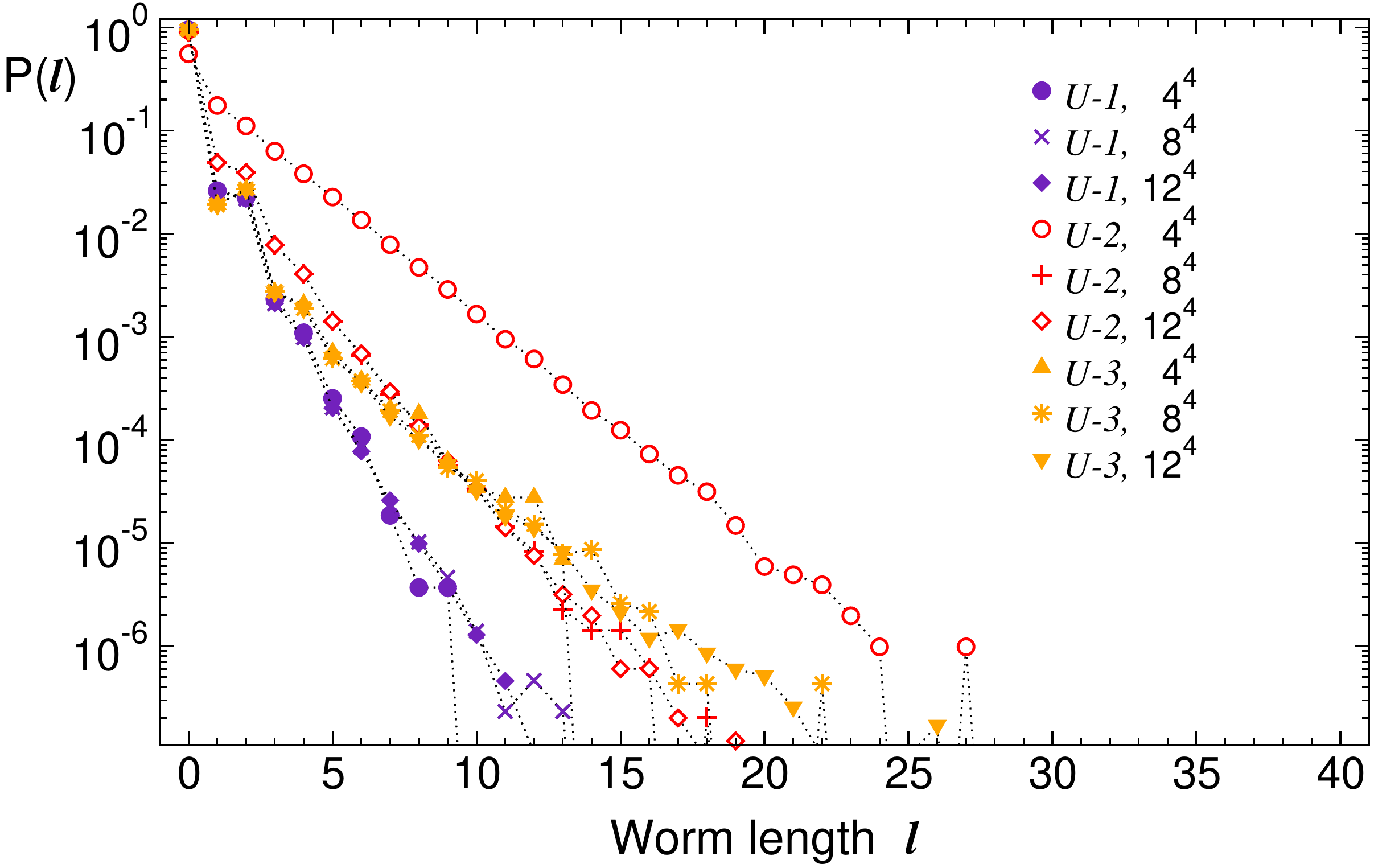}
\end{center}

\caption{Normalized histograms of the worm length for 
the Z$_3$ model (upper plot) 
and the U(1) model (lower plot).}
\label{length}
\end{figure}

Finally, comparing $R_{SWA}$ and $S_{SWA}$ for both the Z$_3$ and U(1) cases, we observe that even though
many worms start successfully, not all of them create non-trivial changes, i.e., 
there is a sizable 
probability that in the second step a worm reverts its initial step. This is also reflected in 
Fig.~\ref{length} where we show the abundance distribution of the worms as a function of their length $l$ defined as the number of segments of a worm.
The distribution decreases roughly exponentially with $l$.  
However, as we shall see in the next subsection, a few long
worms are enough to have a very efficient sampling.

\subsection{Autocorrelation times}
In this subsection we analyze the integrated autocorrelation time  $\tau_{int}^O$ of
several observables $O$ in both models. Since we are comparing two different
algorithms we normalize the autocorrelation times as in \cite{effective}:   
define one sweep as $\tau_0 = 6V_4/{\cal P}$ configurations,  i.e., the average
number of attempts needed to change every  plaquette of the lattice as the unit for
the integrated autocorrelation times $\tau_{int}^O$. In units of updates we have
$\tau_0 = 6V_4/({\cal P} \, N_{updates})$,  where $N_{updates}$ is defined as either
$V_4$ worms for the SWA  or $6V_4$ plaquette updates plus $4V_4$ cube
updates for the LMA, i.e., a total of $10V_4$ local updates.

In order to obtain a measure for the computational effort, the results are multiplied by
the cost ratio ${\cal C}$. In other words we show $\overline{\tau}_{int} = 
{\cal C} \, \tau_{int}/\tau_0$, where $\tau_{int}$ simply is the unnormalized
autocorrelation time in units of updates.  
The statistical errors of autocorrelation times were estimated with a jackknife
analysis and were found at the 10 percent level for the statistics
at our disposal. This is sufficient for the subsequent comparison 
of the two algorithms.

For the autocorrelation analysis we use the same sets and volumes as for the 
discussion of the characteristic quantities of the SWA and the LMA in the previous subsection.
Table \ref{auto-z3w} shows the autocorrelation times in the Z$_3$ case
for the SWA and Table \ref{auto-z3m} is for the LMA.
Similarly, Tables \ref{auto-u1w} and \ref{auto-u1m} correspond to the U(1)
case.

\begin{table}[t!]
\begin{center}
\hspace*{-8mm}
\begin{tabular}{lcccccc}
\hline
Parameters & $V$ &  $\overline{\tau}^{U}_{int}$  & $\overline{\tau}^{\chi_U}_{int}$  & 
$\overline{\tau}^{n}_{int}$  & $\overline{\tau}^{\chi_n}_{int}$  \\
\hline
{\bf Set:} {\it Z-1} & $ 4^3\times 50$ & 180 & 97 & 0.9 & 0.6 \\
$\kappa = 0.1,$      & $ 8^3\times 50$ & 200 & 90 & 1.0 & 0.6 \\
$\beta = 0.6,$       & $12^3\times 50$ & 200 & 92 & 1.0 & 0.6 \\
$\mu = 2.0$          & $16^3\times 50$ & 200 & 88 & 1.3 & 0.8 \\
\hline
{\bf Set:} {\it Z-2} & $ 4^3\times 50$ & 81 & 36 & 25 & 13 \\
$\kappa = 0.1,$      & $ 8^3\times 50$ & 84 & 32 & 25 & 14 \\
$\beta = 0.8,$       & $12^3\times 50$ & $>$ 83 & 38 & 27 & 12 \\
$\mu = 1.6$          & $16^3\times 50$ & $>$ 90 & 37 & 30 & 13 \\
\hline
{\bf Set:} {\it Z-3} & $ 4^4$ & 2.5 & 1.3 & 0.3 & 0.2 \\
$\kappa = 0.5,$      & $ 8^4$ & 5.4 & 2.9 & 0.6 & 0.4 \\
$\beta = 0.28,$      & $12^4$ & 6.0 & 3.1 & 0.6 & 0.5 \\
$\mu = 0.0$          & $16^4$ & 7.7 & 3.2 & 0.7 & 0.5 \\
\hline
\end{tabular}
\end{center}
\caption{Z$_3$ model: SWA autocorrelation times  
for different parameter sets.}
\label{auto-z3w}
\vskip10mm
\begin{center}
\hspace*{-8mm}
\begin{tabular}{lcccccc}
\hline
Parameters & $V$ &  $\overline{\tau}^{U}_{int}$  & $\overline{\tau}^{\chi_U}_{int}$  & 
$\overline{\tau}^{n}_{int}$  & $\overline{\tau}^{\chi_n}_{int}$  \\
\hline
{\bf Set:} {\it Z-1} & $ 4^3\times 50$ & 5400 & 2900 & 5600 & 3100 \\
$\kappa = 0.1,$      & $ 8^3\times 50$ & 5800 & 3000 & 5900 & 3200 \\
$\beta = 0.6,$       & $12^3\times 50$ & 5400 & 3000 & 6100 & 4200 \\
$\mu = 2.0$          & $16^3\times 50$ & 5400 & 3000 & $>$7800 & 4300 \\
\hline
{\bf Set:} {\it Z-2} & $ 4^3\times 50$ & 67 & 48 & 750 & 310 \\
$\kappa = 0.1,$      & $ 8^3\times 50$ & 68 & 51 & 760 & 300 \\
$\beta = 0.8,$       & $12^3\times 50$ & 70 & 49 & 600 & 350 \\
$\mu = 1.6$          & $16^3\times 50$ & 71 & 46 & 600 & 340 \\
\hline
{\bf Set:} {\it Z-3} & $ 4^4$ & 110 & 55 & 59 & 23 \\
$\kappa = 0.5,$      & $ 8^4$ & 110 & 66 & 65 & 24 \\
$\beta = 0.28,$      & $12^4$ & 120 & 69 & 67 & 25 \\
$\mu = 0.0$          & $16^4$ & 130 & 73 & 67 & 27 \\
\hline
\end{tabular}
\end{center}
\caption{Z$_3$ model: LMA autocorrelation times for different parameter sets.}
\label{auto-z3m}
\end{table}

First, we observe that the autocorrelation times for the set close to the  first
order transition (set {\it U-2}) increase with the volume, while the others are
essentially volume independent.  It is also interesting to look at the sets {\it
Z-2} and {\it U-3},  where ${\cal P}_{LMA}$ approaches 6 (see Tables 1 and 2), i.e.,
the configuration space is dominated by closed surfaces, since boundary flux is
costly for these parameter sets.  On the one hand, $\overline{\tau}_{int}^{U}$ and 
$\overline{\tau}_{int}^{\chi_U}$  are larger for the worm algorithm, which is due to
the fact that the worm updates links in every move, so if the Boltzmann weight of the
link variables is very low then  most of the worms have only a few segments (see
Fig.~\ref{length}). On the other hand $\overline{\tau}_{int}$ of the observables
that depend only on the  link occupation number is much smaller for the SWA, a fact which
reflects the very low acceptance rate of the plaquette update of the LMA.

\begin{table}[t!]
\begin{center}
\hspace*{-8mm}
\begin{tabular}{lcccccc}
\hline
Parameters & $V$ &  $\overline{\tau}^{U}_{int}$  & $\overline{\tau}^{\chi_U}_{int}$  & 
$\overline{\tau}^{|\phi|^2}_{int}$  & $\overline{\tau}^{\chi_{|\phi|^2}}_{int}$  \\
\hline
{\bf Set:} {\it U-1}    & $ 4^4$ & 2.2 & 3.1 & 0.6 & 0.3 \\
$\kappa = 5,$    & $ 8^4$ & 2.3 & 1.6 & 0.5 & 0.3 \\
$\lambda = 1,$   & $12^4$ & 2.4 & 1.5 & 0.6 & 0.3 \\
$\beta = 0.40$   & $16^4$ & 2.6 & 1.1 & 0.5 & 0.4 \\
\hline
{\bf Set:} {\it U-2}    & $ 4^4$ & 5.7 & 3.5 & 9.5 & 3.9 \\
$\kappa = 5,$    & $ 8^4$ & 12  & 6.9 & 2.9 & 1.2 \\
$\lambda = 1,$   & $12^4$ & 19  & 7.8 & 3.3 & 1.4 \\
$\beta = 0.65$   & $16^4$ & 21  & 7.9 & 3.1 & 1.6 \\
\hline
{\bf Set:} {\it U-3}    & $ 4^4$ & 1600 & 870 & 1.1 & 0.9 \\
$\kappa = 8,$    & $ 8^4$ & 1700 & 840 & 1.2 & 1.0 \\
$\lambda = 1,$   & $12^4$ & $>$1600 & 740 & 1.8 & 0.9 \\
$\beta = 1.10$   & $16^4$ & $>$1700 & 800 & 2.2 & 1.1 \\
\hline
\end{tabular}
\end{center}
\caption{U(1) model: SWA autocorrelation times for
different parameters.  We attribute the $V = 4^4$ 
value $\overline{\tau}^{\chi_U}_{int} = 3.1$,
which is slightly higher than naively expected, to a finite volume effect
caused by closed surfaces that wind around the rather short
(4 lattice points) compact directions.}
\label{auto-u1w}
\vskip10mm
\begin{center}
\hspace*{-8mm}
\begin{tabular}{lcccccc}
\hline
Parameters & $V$ &  $\overline{\tau}^{U}_{int}$  & $\overline{\tau}^{\chi_U}_{int}$  & 
$\overline{\tau}^{|\phi|^2}_{int}$  & $\overline{\tau}^{\chi_{|\phi|^2}}_{int}$  \\
\hline
{\bf Set:} {\it U-1}    & $ 4^4$ & 5800 & 2900 & 8100 & 4500 \\
$\kappa = 5,$    & $ 8^4$ & 5800 & 3000 & 8600 & 4500 \\
$\lambda = 1.0,$ & $12^4$ & 6100 & 4100 & 7400 & 5000 \\
$\beta = 0.4$    & $16^4$ & 6200 & 4200 & 9100 & 5000 \\
\hline
{\bf Set:} {\it U-2}    & $ 4^4$ & 71   & 48   & 180  & 93   \\
$\kappa = 5,$    & $ 8^4$ & 4700 & 2600 & 7100 & 4100 \\
$\lambda = 1.0,$ & $12^4$ & 7200 & 2800 & 8700 & 4300 \\
$\beta = 0.65$   & $16^4$ & 7300 & 2800 & 9400 & 5000 \\
\hline
{\bf Set:} {\it U-3}    & $ 4^4$ & 460 & 280 & 440 & 300 \\
$\kappa = 8,$    & $ 8^4$ & 430 & 300 & 480 & 300 \\
$\lambda = 1,$   & $12^4$ & 690 & 290 & 450 & 270 \\
$\beta = 1.10$   & $16^4$ & 710 & 280 & 490 & 270 \\
\hline
\end{tabular}
\end{center}
\caption{U(1) model: LMA autocorrelation times for 
different parameters.}
\label{auto-u1m}
\end{table}

In general, comparing the results of both algorithms for the two different models,
we  can conclude that the SWA outperforms  the LMA for a large range of
parameters.   Only in the region of the space of couplings where  the link weight is
very large the worm algorithm has difficulties to sample the system efficiently, a
problem which can easily be overcome with extra cube sweeps or  by adding a worm
with only plaquettes suggested in \cite{endres}.

\section{Summary}
In this article we present a generalization of the worm algorithm to
systems that are described by surfaces with boundaries of flux, i.e., abelian 
Gauge-Higgs systems. Rewriting the standard form of abelian Gauge-Higgs systems
in terms of surfaces and fluxes (dual representation) overcomes the complex action problem at finite
chemical potential. We study Gauge-Higgs systems with two gauge groups Z$_3$ and U(1).
For the Z$_3$ case a chemical potential can be coupled and the system has a complex
action problem.

The key idea of our newly developed surface worm algorithm (SWA)
is to build up filament-like structures where the dual degrees of freedom are changed 
by adding segments built from plaquette variables and two lines of matter flux. 
We compare the SWA to a local Metropolis algorithm (LMA) for the dual representation
and in the cases without a sign problem also to a conventional Monte Carlo simulation
in the standard approach. The comparison is used to establish the correctness of
the SWA in several simulations at different parameter values. 

To study the performance of the SWA we analyze characteristic quantities: the starting
probability, the number of updated plaquettes and the cost efficiency. Based on these
characteristic quantities we conclude that for both gauge
groups and most parameter values the SWA is considerably more efficient than the 
LMA. This finding is confirmed by an analysis of autocorrelation times   
where again the SWA is found to decorrelate faster (partly considerably faster) 
than the LMA.

We expect that the generalization of the worm concept to surface-type degrees of
freedom will contribute to developing new tools for systems with gauge interactions in
a dual language. Another important aspect is that models where the complex action
problem is solved may serve as reference systems for testing other approaches such as
various reweighting and expansion techniques.

\section*{Acknowledgments} 
\vspace{-1mm}
\noindent
We thank Hans Gerd Evertz 
for numerous discussions that helped to shape this project and for 
providing us with the software to compute the autocorrelation times. 
This work was supported by the Austrian Science Fund, 
FWF, DK {\it Hadrons in Vacuum, Nuclei, and Stars} (FWF DK W1203-N16)
and by the Research Executive Agency (REA) of the European Union 
under Grant Agreement number PITN-GA-2009-238353 (ITN STRONGnet).
Y.~Delgado thanks the members of the lattice group in Wuppertal, where
part of this work was done, for a stimulating atmosphere.

\newpage
\noindent
\section*{Appendix: Dual representation for the U(1) Gauge-Higgs system}
\vspace{1mm}
\noindent
In this appendix we summarize a brief derivation of the dual representation of the U(1) Gauge-Higgs system we use in this article. The gauge action $S_G$ is given by (\ref{gauge_action}) with U(1) valued 
link variables $U_{x,\nu}$. The action $S_M$ for the matter field is (\ref{action_higgs}). 
 The partition sum $Z$ is obtained by integrating the Boltzmann factor $e^{-S_G - S_M}$ over all field configurations, 
$Z = \int D[U] D[\phi] e^{-S_G - S_M}$. 
For the Higgs field the measure is a product over all lattice points $x$, and we use polar coordinates  $\phi_x = r_x e^{i \theta_x} 
$ for integrating each $\phi_x$ in the complex plane. The U(1) gauge variables $U_{x,\nu} = e^{i\varphi_{x,\nu}}$ at each link are integrated over the unit circle such that the path integral reads
\begin{equation}
Z  = \int\!\! D[U] D[\phi] \, e^{-S_G - S_M} = \left( \prod_{x,\nu}\! \int_{-\pi}^\pi \!\!\frac{d \varphi_{x,\nu}}{2 \pi} \!\!\right) \!\!
\left(\! \prod_{x}\! \int_{-\pi}^\pi\!\! \frac{d \theta_{x,\nu}}{2 \pi}\!\! \int_0^\infty\!\!\! dr_x \, r_x \!\!\right) \! e^{-S_G - S_H}.
\label{Zconvent}
\end{equation}
The normalization with $2 \pi$ will be useful later. 

The first step to obtain the representation of the full partition sum in terms of loops is to consider the Higgs part of the problem.
For that purpose we define the partition sum of the Higgs system in a gauge background as
\begin{equation}
Z_H = \int\!\! D[\phi] e^{-S_M} = \int\!\! D[\phi]\! \left( \prod_{x,\nu} e^{ \phi_x^\star U_{x,\nu} \phi_{x+\widehat{\nu}}} 
e^{\phi_x U_{x,\nu}^\star \phi_{x+\widehat{\nu}}^\star} \! \right) \!\! \left(\prod_x B(|\phi_x|^2)\!\right) ,
\end{equation}
where we have slightly reorganized the nearest neighbor terms and write the corresponding sums in the exponent as 
a product of exponentials. The mass- and $\phi^4$-terms are taken into account in $B(r^2) = \exp(-\kappa r^2 - \lambda r^4)$.

The next step is an expansion of the Boltzmann factors for the nearest neighbor terms (use $U_{x,\nu}^{\;\;\star} = U_{x,\nu}^{\;\;-1}$):
\begin{eqnarray}
\hspace*{-6mm}&& \prod_{x,\nu} \exp\!\left(\phi_x^\star U_{x,\nu}    \phi_{x+\widehat{\nu}}\right) 
 \exp\!\left(  \phi_x U_{x,\nu}^\star \phi_{x+\widehat{\nu}}^\star \right)  =   
\\
\hspace*{-6mm}&& 
\sum_{\{ n, \overline{n}\}} \!\!
\left( \prod_{x,\nu}\! \frac{U_{x,\nu}^{\;\;n_{x,\nu}} \, 
{U_{x,\nu}^\star}^{\overline{n}_{x,\nu}}}{n_{x,\nu}! \, \overline{n}_{x,\nu}!} \right) \!
\left( \prod_{x,\nu}\! \Big(\phi_x^\star \phi_{x+\widehat{\nu}}\Big)^{n_{x,\nu}} \, 
\Big(\phi_x \phi_{x+\widehat{\nu}}^\star\Big)^{\overline{n}_{x,\nu}} \! \right) \!= 
\nonumber \\
\hspace*{-6mm}&& 
\sum_{\{ n, \overline{n}\}} \!\!
\left( \prod_{x,\nu}\! \frac{U_{x,\nu}^{\;\;n_{x,\nu}-\overline{n}_{x,\nu}}}{n_{x,\nu}! \, \overline{n}_{x,\nu}!} \right) 
\left(\!  {\phi_x^{\,\star}}^{\sum_\nu [ n_{x,\nu} + \overline{n}_{x-\widehat{\nu},\nu} ] } 
\, {\phi_x}^{\sum_\nu 
[ \overline{n}_{x,\nu} + n_{x-\widehat{\nu},\nu} ] }  \right) ,
\nonumber 
\end{eqnarray}
where the expansion variables $n_{x,\nu}$ and $\overline{n}_{x,\nu}$ are non-negative integers attached to the links of the lattice. 
By $\sum_{\{n,\overline{n}\}}$ we denote the sum over all configurations of the expansion variables $n_{x,\nu}, 
\overline{n}_{x,\nu} \in [0,\infty)$. The partition sum of the Higgs field now reads
\begin{eqnarray}
Z_H &\!\!\! = \!\!\!&\!\! \sum_{\{ n, \overline{n}\}}  \!\!
\left( \prod_{x,\nu}\! \frac{U_{x,\nu}^{\;\;n_{x,\nu}-\overline{n}_{x,\nu}}}{n_{x,\nu}! \, \overline{n}_{x,\nu}!} \right) \!\!
\left(\! \prod_x \int_{-\pi}^\pi \frac{d\theta_x}{2\pi} e^{-i\theta_x \sum_\nu [ n_{x,\nu}  - \overline{n}_{x,\nu} - 
( n_{x-\widehat{\nu},\nu}   - \overline{n}_{x-\widehat{\nu},\nu}) ] }\! \right)  \nonumber \\
& &\hspace{10mm} \times
\left(\prod_x \int_{0}^\infty\!\!\! dr_x \; r_x^{1 + \sum_\nu [ n_{x,\nu} + n_{x-\widehat{\nu},\nu} 
+ \overline{n}_{x,\nu} + \overline{n}_{x-\widehat{\nu},\nu} ] }  \; B\big(r_x^2 \big) \right)\! . 
\end{eqnarray}
The integrals over the phase give rise to Kronecker deltas, which for notational convenience here
we write as $\delta(n)$. The integrals over the modulus we abbreviate as
\begin{equation}
P(n) \; = \; \int_0^\infty dr \, r^{n+1} \,  B\big(r^2) \; = \; \int_0^\infty dr \, r^{n+1} \,  e^{-\kappa r^2 - \lambda r^4} \; .
\label{ww}
\end{equation}
They can easily be computed numerically. 
The Higgs field partition sum now reads:
\begin{eqnarray}
Z_H \!\!&\!\! = \!\!&\!\! \sum_{\{ n, \overline{n}\}}\!\!  
\left( \prod_{x,\nu}\! \frac{U_{x,\nu}^{\;\;n_{x,\nu}-\overline{n}_{x,\nu}}}{n_{x,\nu}! \, \overline{n}_{x,\nu}!} \right) \!\!
\left(\! \prod_x \delta\!\left( \sum_\nu \big[ n_{x,\nu}  - \overline{n}_{x,\nu} - 
( n_{x-\widehat{\nu},\nu}   - \overline{n}_{x-\widehat{\nu},\nu}) \big] \! \right)\!\right)
\nonumber \\
& \!\!  \!\!& \hspace{10mm} \times
\left(\! \prod_x  P\!\left( \sum_\nu \big[ n_{x,\nu} 
+ \overline{n}_{x,\nu} + n_{x-\widehat{\nu},\nu} +  \overline{n}_{x-\widehat{\nu},\nu} \big]  \right) \right)\! . 
\end{eqnarray}
In this form the Higgs fields are completely eliminated and the partition sum is a sum over  
configurations of the $n$ and $\overline{n}$. The allowed configurations of the $n$ and $\overline{n}$ are subject to local constraints at each 
site $x$ enforced by the Kronecker deltas, i.e., at each site $x$ the variables must obey  $\sum_\nu [ n_{x,\nu}  - \overline{n}_{x,\nu} - 
( n_{x-\widehat{\nu},\nu}   - \overline{n}_{x-\widehat{\nu},\nu}) ]  = 0$. 

In the current representation the constraints mix both the 
$n$ and the $\overline{n}$ variables. The structure of the constraints can be simplified by introducing new variables
$l_{x,\nu} \in (-\infty,\infty)$ and $\overline{l}_{x,\nu} \in [0,\infty)$. They are related to the old variables by
\begin{equation}
n_{x,\nu} - \overline{n}_{x,\nu} = l_{x,\nu} \qquad
\mbox{and} \qquad  n_{x,\nu} + \overline{n}_{x,\nu} = |l_{x,\nu}| + 2 \overline{l}_{x,\nu} \; ,
\label{newintegers}
\end{equation} 
and the sum over all configurations of the $n,\overline{n}$ variables 
can be replaced by a sum over $l$- and $\overline{l}$-configurations.
The partition sum turns into
\begin{eqnarray}
Z_H &\!\! = \!\!& \sum_{\{ l, \overline{l}\}}  \left( \prod_{x,\nu}\! \frac{U_{x,\nu}^{\;\;l_{x,\nu}}  }{(|l_{x,\nu}| + \overline{l}_{x,\nu})! \, \overline{l}_{x,\nu}!} \right) \!\!
\left(\! \prod_x \delta\left( \sum_\nu \big[ l_{x,\nu}  -  l_{x-\widehat{\nu},\nu}  \big] \! \right) \right)
\nonumber \\
& \!\!  \!\!& \hspace{10mm} \times
\left(\!\prod_x
P\!\!\left( \sum_\nu \big[ |l_{x,\nu}| +  |l_{x-\widehat{\nu},\nu}| + 2( \overline{l}_{x,\nu} + \overline{l}_{x-\widehat{\nu},\nu}) \big]  \right) \right)\! . 
\label{ZHfinal}
\end{eqnarray}
In the final form (\ref{ZHfinal}) of the Higgs field partition sum, 
which we now refer to as dual representation, the constraints no longer mix the two types of flux variables. 
Obviously only the $l$-fluxes are subject to conservation of flux at each site $x$, i.e., only they must obey 
$\sum_\nu [ l_{x,\nu} - l_{x-\widehat{\nu},\nu}] = 0$ for all $x$.

Having mapped the Higgs field partition sum to the flux form  (\ref{ZHfinal}) we now apply similar steps to the gauge fields to obtain the dual 
representation of the full partition sum (\ref{Zconvent}). We write the full partition sum as $Z = \int D[U] e^{-S_G} Z_H$ and find
\begin{eqnarray}
Z &=& \sum_{\{ l, \overline{l}\}}  \left( \prod_{x,\nu}\! \frac{ 1  }{(|l_{x,\nu}| + \overline{l}_{x,\nu})! \, \overline{l}_{x,\nu}!} \right) \!\!
\left(\! \prod_x \delta\left( \sum_\nu \big[ l_{x,\nu}  -  l_{x-\widehat{\nu},\nu}  \big] \! \right)\! \right)
\nonumber
\\
&& \hspace{-3mm} \times
\left(\! \prod_x P\!\left( \sum_\nu \big[ |l_{x,\nu}| +  |l_{x-\widehat{\nu},\nu}| + 2( \overline{l}_{x,\nu} + \overline{l}_{x-\widehat{\nu},\nu}) \big]\!  \right)\! \right)
Z_G[l] \,,
\label{Zfull1}
\end{eqnarray}
where we have interchanged the sum over the flux configurations and the integral over the gauge fields.
The gauge field partition sum with link insertions according to a flux configuration $l$ is defined as
\begin{equation}
Z_G[l] = \int D[U] e^{-S_G} 
\, \prod_{x,\nu}\!  U_{x,\nu}^{\;\;l_{x,\nu}}  \; .
\label{ZGK}
\end{equation}
The gauge action $S_G$ as defined in (\ref{gauge_action}) is a sum over plaquettes. 
We thus may write the Boltzmann factor $e^{-S_G}$ as a 
product over plaquettes and, as done for the Higgs field, we expand the corresponding exponentials into power series:
\begin{eqnarray}
\prod_{x,\sigma < \tau}\!\! e^{\frac{\beta}{2} U_{x,\sigma\tau}} \, e^{\frac{\beta}{2} U_{x,\sigma\tau}^\star } &\!\!\!=\!\!\!& 
\sum_{\{ m, \overline{m}\}} \! \left(
\prod_{x,\sigma < \tau }\! \frac{(\frac{\beta}{2})^{m_{x,\sigma\tau} + \overline{m}_{x,\sigma\tau}}}{m_{x,\sigma\tau}! \, \overline{m}_{x,\sigma\tau}!}
 {U_{x,\sigma\tau}}^{m_{x,\sigma\tau}} \; {U_{x,\sigma\tau}^\star}^{\overline{m}_{x,\sigma\tau}} \right)
\nonumber \\
&\!\!\!=\!\!\!& 
\sum_{\{ m, \overline{m}\}}  \!
\left( \prod_{x,\sigma < \tau }\! \frac{(\frac{\beta}{2})^{m_{x,\sigma\tau} + 
\overline{m}_{x,\sigma\tau}}}{m_{x,\sigma\tau}! \, \overline{m}_{x,\sigma\tau}!} \, \right) 
\label{gaugeboltz}
\\
&\!\!\!\times\!\!\!&  \left( \prod_{x,\nu} {\,U_{x,\nu}}^{\sum_{\nu < \alpha} \big[ p_{x,\nu\alpha} - p_{x-\widehat{\alpha},\nu\alpha}\big]
- \sum_{\alpha<\nu} \big[ p_{x,\alpha\nu} - p_{x-\widehat{\alpha},\alpha\nu}\big] }\! \right) .
\nonumber
\end{eqnarray}
We introduced the expansion variables $m_{x,\sigma\tau}, \overline{m}_{x,\sigma\tau} \in
[0,\infty)$ attached to the plaquettes, and  by
$\sum_{\{m,\overline{m}\}}$ we denote the sum over all configurations of the  expansion variables.
In the second step we inserted  the explicit expressions for the plaquettes in terms of the link
variables, i.e., $U_{x,\sigma\tau} = U_{x,\sigma} U_{x+\widehat{\sigma}, \tau}
U_{x+\widehat{\tau},\sigma}^\star U_{x,\tau}^\star$, and reorganized the product over powers of
links variables. Here we already introduced  $m_{x,\nu\alpha} - \overline{m}_{x,\nu\alpha}  =
p_{x,\nu\alpha} $.  This combination of the expansion variables plays the same role as the
transformutation (\ref{newintegers}) used in the Higgs case for the simplification of the
constraints. Exactly the same step is now implemented here: We promote $p_{x,\nu\alpha}  \in
(-\infty,\infty)$ into new dynamical variables, which  together with another set of variables,
$q_{x,\nu\alpha} \in [0,\infty)$, gives the final set of variables we use for the gauge fields. 
The $p$ and $q$ variables are related to the $m$ and  $\overline{m}$ variables via (compare
(\ref{newintegers}))
\begin{equation}
m_{x,\nu\alpha} - \overline{m}_{x,\nu\alpha} = p_{x,\nu\alpha} \quad \;
\mbox{and} \quad \; m_{x,\nu\alpha} + \overline{m}_{x,\nu\alpha} = |p_{x,\nu\alpha}| + 2 q_{x,\nu\alpha} \; .
\label{newintegers2}
\end{equation} 
We will refer to the variables $p$ as plaquette occupation numbers or simply plaquette variables. 
Using the new variables (\ref{newintegers2})  and inserting the expanded Boltzmann factor (\ref{gaugeboltz}) back into (\ref{ZGK}) we find
\begin{eqnarray}
Z_G[l] &\!\! =\!\! & \sum_{\{p,q\}} \left( \prod_{x,\sigma < \tau} \frac{(\frac{\beta}{2})^{|p_{x,\sigma\tau}| + 2 q_{x,\sigma\tau}}}{(
|p_{x,\sigma\tau}| + q_{x,\sigma\tau})! \, q_{x,\sigma\tau}!} \right)
\\
&\!\!\times\!\! & \!\!\!\!\left(\!\prod_{x,\nu} \int_{-\pi}^\pi \!\!\frac{d \varphi_{x,\nu}}{2\pi} e^{i\varphi_{x,\nu} \left( \sum_{\nu < \alpha} \!\big[ p_{x,\nu\alpha} - p_{x-
\widehat{\alpha},\nu\alpha}\big]
- \sum_{\alpha<\nu}\! \big[ p_{x,\alpha\nu} - p_{x-\widehat{\alpha},\alpha\nu}\big]  + l_{x,\nu} \right) }\!\!\right)\!\! .
\nonumber 
\end{eqnarray}
The integrals in the last product are again representations of Kronecker deltas and give rise to constraints that are located at the links of the lattice. The summations over the variables $q_{x,\sigma\tau}$ can be done in closed form using the well known series representation of the modified Bessel functions
\begin{equation}
\sum_{q=0}^\infty \frac{(\frac{\beta}{2})^{|p|+2q}}{(|p|+q)!\, q!} \; = \; I_{|p|}(\beta) \; = \; I_{p}(\beta) \; ,
\label{bessel}
\end{equation}
where in the last step we used the fact that the modified Bessel functions $I_n(z)$ are even in their index $n$.
Thus we finally end up with the following representation for the gauge field partition sum
\begin{eqnarray}
Z_G[l] &\!\! =\!\! & \sum_{\{p\}} \left( \prod_{x,\sigma < \tau} I_{p_{x,\sigma\tau}}(\beta) \right) 
\\
&\!\!\times\!\!&
\left( \prod_{x,\nu} \delta\left(\sum_{\nu < \alpha} \big[ p_{x,\nu\alpha} - p_{x-\widehat{\alpha},\nu\alpha}\big]
- \sum_{\alpha<\nu} \big[ p_{x,\alpha\nu} - p_{x-\widehat{\alpha},\alpha\nu}\big]  + l_{x,\nu} \right) \right).
\nonumber
\end{eqnarray}
Putting this back into the full partition sum (\ref{Zfull1}) we obtain the final result for the dual representation of the partition sum for the U(1) 
Gauge-Higgs model as given in Eqs.~(\ref{u1dual}), (\ref{weight_u1}) and (\ref{constraint_u1}).

Let us finally comment on the possibility to couple chemical potential $\mu$: The derivation of the dual representation remains essentially the same, with additional factors $e^{\pm \mu}$ for the temporal links. In the final expression these factors give different (real and positive) weight for positive and negative temporal $l$-flux. In this paper we only consider one flavor of the Higgs field, and Gauss law does not allow to construct configurations that obey all constraints at $\mu > 0$. In an upcoming study \cite{dualu1_ref} we will present results for two flavors of oppositely charged Higgs fields, where non-zero chemical potential is possible and interesting condensation phenomena can be studied.

\end{document}